\documentclass[manuscript]{aastex}
\usepackage{amssymb,amsmath}

\shorttitle{Wavelet Analyses of Solar Wind Composition Variability}
\shortauthors{Edmondson et al.}

\begin{document}

\title{Analysis of High Cadence In-Situ Solar Wind Ionic Composition Data Using Wavelet Power Spectra Confidence Levels}

\author{J. K. Edmondson\altaffilmark{1}, B. J. Lynch\altaffilmark{2}, S. T. Lepri\altaffilmark{1}, and T. H. Zurbuchen\altaffilmark{1}}
\affil{\altaffilmark{1}Department of Atmospheric, Oceanic, and Space Science, University of Michigan, Ann Arbor, MI 48109}
\affil{\altaffilmark{2}Space Sciences Laboratory, University of California, Berkeley, CA 94720}

\begin{abstract}

The variability inherent in solar wind composition has implications for the variability of the physical conditions in its coronal source regions, providing constraints on models of coronal heating and solar wind generation. We present a generalized prescription for constructing a wavelet power significance measure (confidence level) for the purpose of characterizing the effects of missing data in high cadence solar wind ionic composition measurements. We describe the data gaps present in the 12-minute ACE/SWICS observations of ${\rm O}^{7+}/{\rm O}^{6+}$ during 2008. The decomposition of the in-situ observations into a `good measurement' and a `no measurement' signal allows us to evaluate the performance of a filler signal, i.e., various prescriptions for filling the data gaps. We construct Monte Carlo simulations of synthetic ${\rm O}^{7+}/{\rm O}^{6+}$ composition data and impose the actual data gaps that exist in the observations in order to investigate two different filler signals: one, a linear interpolation between neighboring good data points, and two, the constant mean value of the measured data. Applied to these synthetic data plus filler signal combinations, we quantify the ability of the power spectra significance level procedure to reproduce the ensemble-averaged time-integrated wavelet power per scale of an ideal case, i.e. the synthetic data without imposed data gaps. Finally, we present the wavelet power spectra for the ${\rm O}^{7+}/{\rm O}^{6+}$ data using the confidence levels derived from both the Mean Value and Linear Interpolation data gap filling signals and discuss the results.

\end{abstract}


\section{Introduction}

Decades of in-situ plasma observations have revealed a rich picture of the solar wind \citep[][and references therein]{Zurbuchen07}, whose overall structure and magnetic topology follows the solar magnetic activity cycle. Heliospheric solar wind observations reflect the structure of their coronal source regions: a relatively cool, fast solar wind with relatively homogeneous ionic composition and elemental abundances originating from coronal holes \citep{Geiss95,McComas02}, and a relatively hot, slow solar wind that exhibits considerably more variability in ionic composition and elemental abundances, originating either directly from within the vicinity of coronal streamers \citep{Gosling97,Zurbuchen02}. In-situ observations of ionic charge state composition, especially of carbon (${\rm C}^{6+}/{\rm C}^{4+}$) and oxygen (${\rm O}^{7+}/{\rm O}^{6+}$) offer insight into coronal dynamics at temperatures of order one million degrees \citep[e.g.,][]{vonSteiger1997,Zhao09,Landi12,Gilbert12}. Identifiable temporal scales from within the compositional variability may provide insights into the nature of the source regions of the solar wind.

Wavelet transforms are used to identify transient structure coherency as well as global periodicities in time series data \citep[see e.g.,][]{Daubechies1992, TorrenceCompo1998, Liu2007}. Wavelet analyses have an advantage over traditional spectral methods by being able to isolate both large timescale and small timescale periodic behavior that occur over only a subset of the time series. Thus, we are able to analyze the frequency decomposition as a function of time. This is extremely useful if we expect the time series to originate from either a time varying source region or, equivalently, to be consecutively sampling many different source regions with varying physical properties, such as in the solar wind.  

Recently, \citet{Katsavrias12} used wavelets to examine four solar cycles worth of solar wind plasma, interplanetary magnetic field, and geomagnetic indices to verify intermittent periodicities on timescales shorter than the solar cycle. Common solar timescales ranging from a decade down to hours have been characterized, and timescales of the order of a Carrington Rotation period (approx. 27 days) and shorter (e.g., 14, 9, and 7 days) have been consistently identified in various heliospheric and geomagnetic data \citep[e.g.,][]{Bolzan05,Fenimore78,Gonzalez87,Gonzalez93,Mursula98,Nayar01,Nayar02,Svalgaard75}. \citet{Temmer07} linked the 9 day timescale to coronal hole variability in the declining phase of solar cycle 23 and \citet{Neugebauer97} used wavelet analyses of {\it Ulysses} solar wind speed data to investigate polar microstreams occurring on timescales of 16 hours. 

Wavelet power spectra are a powerful tool to identify and characterize structures with specific transient timescales and global periodicities, but all commonly used algorithms require fully populated data-sets. That is inconsistent with solar wind composition data -- as well as almost all in situ data-sets -- because data gaps occur for a number of reasons. The experiment may undergo maintenance and data may not be available, or the signal to noise of the instrument at a given time may have prevented a valid and accurate measurement. Thus, care must be taken to account for spurious results caused by such data gaps. Thus, to identify characteristic timescales smaller than the largest gap duration, one must either break-up the full data set into disjoint segments of continuous data measurements, or quantify the spurious information introduced into the data set by filling-in the no-measurement times. It is with the latter solution that the methodology described in this paper is concerned.

Our purpose here is to describe a generalized procedure for the construction of wavelet power significance levels that quantify the relative influence of a filler signal of generally arbitrary form interleaved within a measured data signal. The decomposition of the time series allows for a similar decomposition of the total wavelet power spectrum, and thereby quantifying the power spectra associated with the filler signal and nonlinear interference, for comparison against the measured data signal power. Using the decomposition of the signal power spectra, we identify a statistical confidence level against the null hypothesis that a given feature in the total wavelet power spectrum is due to the filler signal and/or interference effects; in other words, we construct a significance measure for the the total wavelet power spectrum that identifies power spectral features resulting from the measured signal.

The structure of the paper is as follows. In Section~\ref{S:WaveletCharacteristics} we briefly review the wavelet transform, power spectrum, and methods for identifying global periodicities (akin to Fourier modes) as well as transient coherency characteristics. In Section~\ref{S:Data} we discuss the solar wind ionic composition data obtained by ACE/SWICS during the quiet solar conditions of 2008, and the origin and characteristics of no-measurement data gaps in the context of wavelet analysis. In Section~\ref{S:DataReductionScheme} we derive the wavelet power statistical confidence level to characterize the effects, and quantify the influence of no-measurement gaps in the data. In Section~\ref{S:MonteCarlo} we evaluate the performance of two filler signal forms (Linear Interpolation and constant Mean Value) using ensemble-averaged Monte Carlo simulations of a statistical ${\rm O}^{7+}/{\rm O}^{6+}$ ratio model random (1$^{st}$-order Markov) process. In Section~\ref{S:WaveletO7O6} we examine the wavelet power spectra of actual 12 minute ${\rm O}^{7+}/{\rm O}^{6+}$ data from 2008 with the Linear Interpolation filler signal for the high cadence data gaps, and present our conclusions in Section~\ref{S:Conclusions}.

\section{Rectified Wavelet Power Spectrum Analysis}
\label{S:WaveletCharacteristics}

The wavelet transform of a time series $T(t)$ is given by
\begin{equation} \label{E:WaveletTransform}
W_{\rm T}( t , s ) = \int {\rm T} ( t' ) \ \psi^* ( t' , t , s ) \ dt'.
\end{equation}
\noindent In our calculations, the wavelet bases are generated from the Morlet family, though we note all following analysis is valid for any wavelet basis family. The Morlet family is a time-shifted, time-scaled, complex exponential modulated by a Gaussian envelope,
\begin{equation}
\psi \left( t' , t , s \right) = \frac{\pi^{1/4}}{\vert s \vert^{1/2}} {\rm exp}\left[ i \omega_{0} \left( \frac{t' - t}{s} \right) \right] {\rm exp}\left[ - \frac{1}{2} \left( \frac{t' - t}{s} \right)^2 \right]
\end{equation}
\noindent where $ \left( t' , t \right) \in I_{T} \times I_{T} \subset \mathbb{R} \times \mathbb{R}$ is the time and time-translation center, respectively, and $s \in I_{S} \subset \mathbb{R}$ is the timescale over which the Gaussian envelope is substantially different from zero. The $\omega_{0} \in \mathbb{R}$ is a non-dimensional frequency parameter defining the number of oscillations of the complex exponential within the Gaussian envelope; we set $\omega_{0} = 6$, yielding approximately three oscillations within the envelope.

The wavelet power spectrum is given by, $\vert W_{\rm T}( t , s ) \vert^2$, for $\psi , T \in L^2 \left( \mathbb{R} \right)$. \citet{TorrenceCompo1998} identify a bias in favor of large timescale features in the canonical power spectrum, which they attribute to the width of the wavelet filter in frequency-space; at large timescales the function is highly compressed yielding sharper peaks of higher amplitude. Equivalently, high frequency peaks tend to be underestimated because the wavelet filter is broad at small timescales. \citet{Liu2007} showed this effect is the difference between the energy and the integration of the energy with respect to time, and thus may be rectified, in practice, by multiplying the wavelet power spectra by the corresponding frequency. Thus, throughout this paper we use the rectified power spectrum given by 
\begin{equation}
\mathcal{P}_{T} \left( t , s \right) = \vert s \vert^{-1} \vert W_{T} \left( t , s \right) \vert^2.
\end{equation}

The (rectified) wavelet power spectrum for a general time series and wavelet basis can be highly structured and complex. For solar wind composition data, the time series will likely include characteristic global solar oscillation frequencies, such as the approximate 27-day solar rotation period. In addition to any characteristic global oscillations, the time series will likely be full of transient (non-stationary) `coherent structures' in which variations in the coronal source parameters lead to local variations in the composition ratio data. Thus, we define a localized `coherent structure' as data points that become locally elevated with respect the surrounding measurements, which we note, may spread over a variety of timescales.

Global periodicities feature as horizontal bands of (relatively) strong power in the wavelet power spectrum. We characterize the global periodic behavior by integrating the wavelet power spectrum along each correlation timescale to calculate the energy contained in all wavelet coefficients at that timescale; this is known as the global wavelet power spectrum \citep[see e.g.,][]{LeWang03, Bolzan05}, and for the Morlet family is akin to the Fourier modes. The global periodic frequencies within the time series (e.g., Fourier modes) are identified with the local maxima in the integrated power per timescale. Transient (non-stationary) `coherent structures' feature as localized 2D maxima in the wavelet power spectrum. The timescale corresponding to such local 2D maxima demarcates the coherency size of the transient feature.

\section{{\it ACE}/SWICS Measurements of ${\rm O}^{7+}/{\rm O}^{6+}$ Solar Wind Composition}
\label{S:Data}

The {\it Advanced Composition Explorer} spacecraft (ACE) is currently in orbit about the L1 point, $\sim$1.5 million km sunward of Earth \citep{Stone1998}. Here we analyze data obtained with the Solar Wind Ionic Composition Spectrometer \citep[SWICS;][]{Gloeckler98}. SWICS is a time-of-flight (TOF) mass spectrometer paired with energy-resolving solid-state detectors (SSDs) and an electrostatic analyzer (ESA) that measures the ionic composition of the solar wind. Ions with the appropriate energy per charge are selected in the ESA. Ion speed is determined in the TOF telescope and the residual energy measured by the SSDs enables particle identification. These measurements allow the independent determination of mass, $M$, charge, $Q$, and energy, $E$, and are virtually free of background contamination \citep[e.g., see][]{vonSteiger2000}. 

In order that the wavelet transform defined by equation (\ref{E:WaveletTransform}) to be well-defined, and the analysis capable of identifying coherent structure and global oscillation frequencies to the highest time resolution, the (full year) input time series of charge state ratio values must be fully populated at the given cadence. Typically the composition measurements in 1- and 2-hour averages have superb counting statistics, however in the highest time resolution data (12 minute cadence) the flux levels are occasionally too low for a valid measurement to be recorded. For an ionic composition ratio, the presence of a valid data point is subject to the relatively restrictive condition that the ACE/SWICS instrument must have made a measurement for both numerator and denominator with enough counting statistics such that the data reduction algorithm derives a value at the given time resolution, and the denominator must be non-zero. In other words, under low-flux conditions, which occurred throughout 2008, the charge state ratios under consideration occasionally could not be constructed.

\begin{figure}
\epsscale{1.0}
\plotone{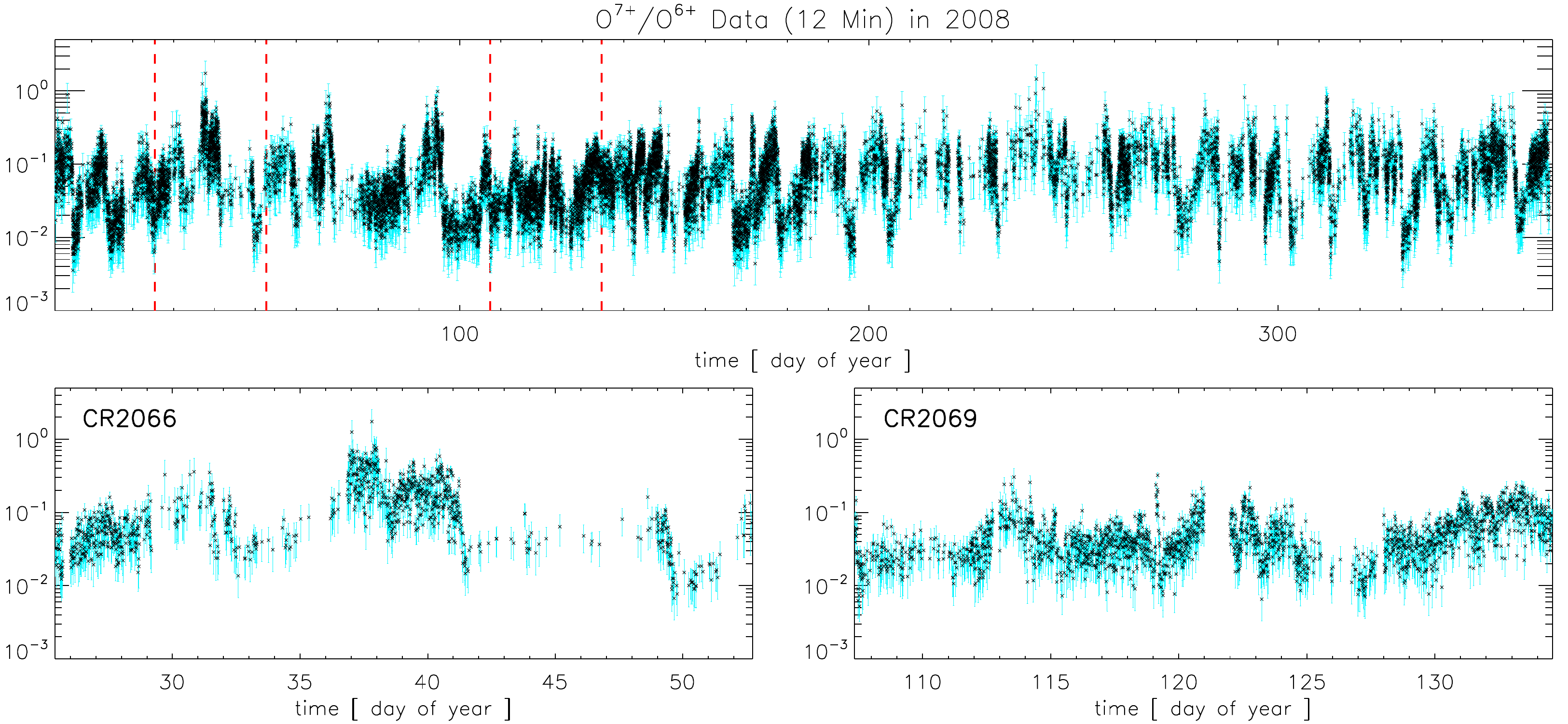}
\caption{Top panel plots the 12-minute ${\rm O}^{7+}/{\rm O}^{6+}$ data for 2008. The vertical dashed lines denote Carrington Rotations 2066 and 2069 which are shown in the bottom panels. The intermittent data gaps are readily visible in the lower panels. 
               \label{datafig}}
\end{figure}

\begin{figure}
\epsscale{0.75}
\plotone{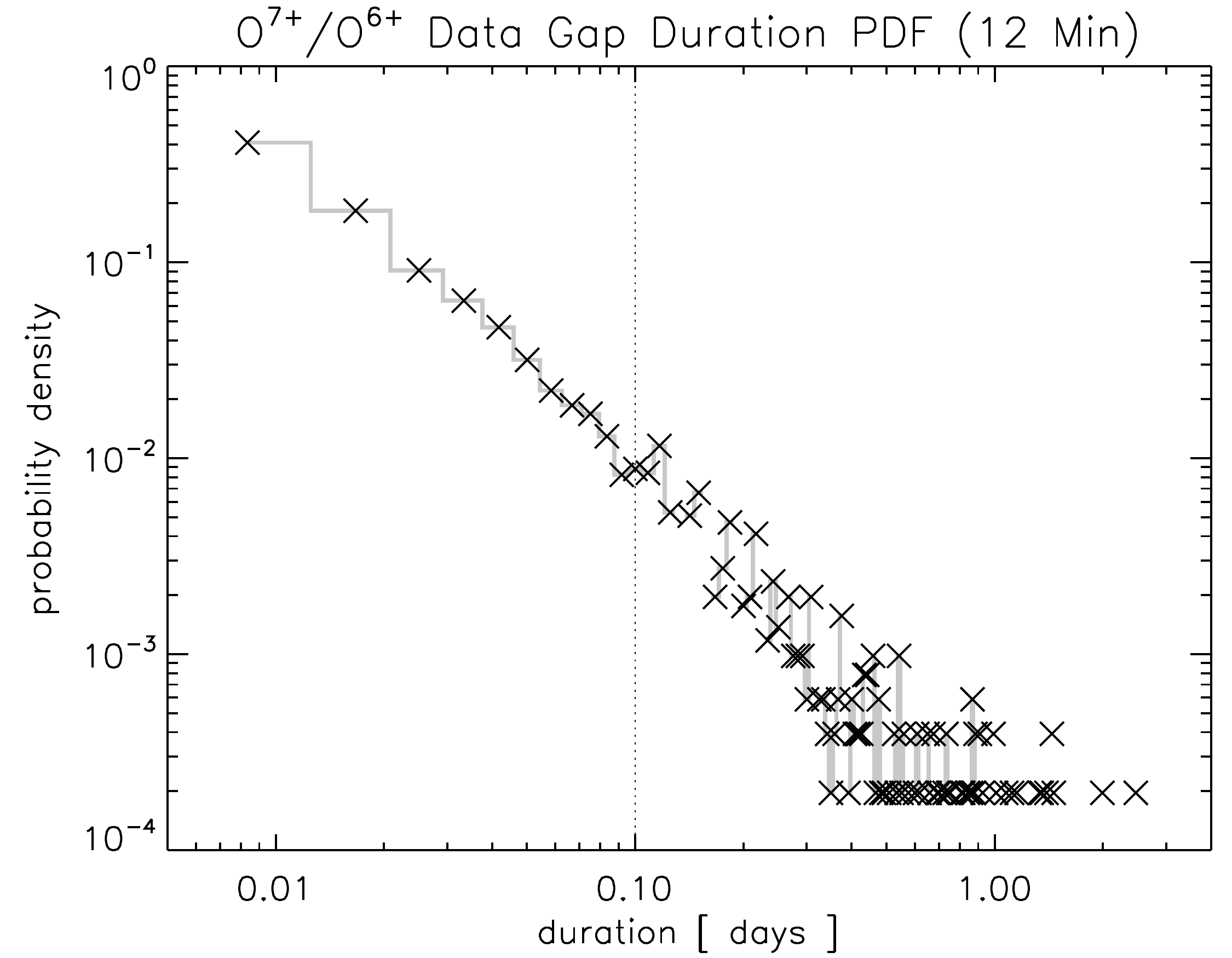}
\caption{Probability density function of the duration of the 12-minute ${\rm O}^{7+}/{\rm O}^{6+}$ data gaps during 2008. Data gaps shorter than 2.4 hours (0.1~day, vertical dotted line) make up 90.4\% of the distribution. 
               \label{MissingDataPDF}}
\end{figure}

The top panel of Figure \ref{datafig} shows 12-minute average ${\rm O}^{7+}/{\rm O}^{6+}$ ACE/SWICS measurements for the full 2008 year. Qualitatively, there are no discernible gaps in the time series over the full year. However, upon closer viewing for example of individual Carrington Rotations (bottom panels of Figure~\ref{datafig}), the ubiquity of such no-measurement data gaps becomes clear. Note the sporadic nature of the gaps between 42 and 49 days within CR2066, as well as the day 121 in CR2069. The frequency of the data gap durations is quantified in Figure~\ref{MissingDataPDF} which plots the probability distribution function (PDF) of the missing data time durations. From this, we find the vast majority, $90.4\%$, of the no-measurement durations occur on timescales less than 0.1 days (2.4 hours). In addition, $9.4\%$ of the gap durations occur on timescales between 0.1 and 1 days. Only $0.2\%$ of the data gaps have durations greater than 1 day. The single maximum no-measurement duration is 2.5 days (occurring at 214 days into the year).

Any analysis of the wavelet power spectrum of a time series that includes data gaps is only valid to the timescales of the largest data gap (in this case, 2.4 days). Below this timescale, one must break the full year data time series into subsets of continuous measurement durations, and perform similar power spectrum analyses on the individual subset time series. Such a procedure, while valid, leads to a host of issues. For example, any physical global oscillations on timescales below the largest data gap are lost. In addition, the boundary effects associated with the cone of influence within the individual wavelet power spectra \citep[see e.g.,][]{TorrenceCompo1998}, become amplified as the size of the data set decreases. In this paper, we take a different approach. We retain as much physical information below the largest data gap timescale as possible by filling the data gaps with a particular signal form and quantifying the propagation of new information introduced into the system throughout the analysis.

\section{Constructing Wavelet Power Confidence Levels to Characterize the Effects of Data Gap}
\label{S:DataReductionScheme}

In order to attempt to keep any physical information of global oscillation frequencies and coherent structures below the timescale of the largest data gap, we require a fully-populated time series for the full time interval under scrutiny. Therefore, we introduce a particular signal form model to fill the data gaps, and quantify the new information introduced to the system by constructing a statistical confidence level as a measure of the influence of the filler signal on the total wavelet power spectrum.

\subsection{Wavelet Power Spectrum from a Superposition of Signals}

From a qualitative standpoint, the wavelet power at any given timescale is determined by several factors, the strength of the measured signal, the strength of the filler signal, and interference effects between the data signal and filler signal. To quantify this decomposition of the wavelet power spectrum, we first note that the full time interval, $t \in I_{T} \subset \ \mathbb{R}$, may be decomposed into (discontinuous) interleaved subsets of measurement time, $t \in I_{D} \subset I_{T} \subset \ \mathbb{R}$, and no-measurement time, $t \in I_{F} \subset I_{T} \subset \ \mathbb{R}$. Note, $I_{T} = I_{D} + I_{F}$.

With this decomposition of the time interval, we may then decompose the full time series, $T(t)$, into a linear superposition of two signals over the full duration consisting of the measurement data signal, $D(t)$, such that the values within the no-measurement intervals are set equal to zero; and the no-measurement filler signal, $F(t)$, in which non-zero values fall within the no-measurement intervals.
\begin{eqnarray}
D \left( t \right) = & \left\{ \begin{array}{ccr} D \left( t \right) & , & t \in I_{D} \\ 0 & , & t \in I_{F}
\end{array} \right. \\
F \left( t \right) = & \left\{ \begin{array}{ccr} F \left( t \right) & , & t \in I_{F} \\ 0 & , & t \in I_{D}
\end{array} \right. 
\label{E:SignalDecomposition}
\end{eqnarray}

The full time series is, therefore, $T(t) = D(t) + F(t) \ \forall \ t \in I_{T}$. We note, the full time series may contain zero values, though only where zero measurements were in fact made. On the other hand, the no-measurement intervals are filled by a model of a known functional form.

To demonstrate the procedure, we construct the following example time series shown in Figure~\ref{O7O6model}. The data signal, $D(t)$, shown in black in the top panel of Figure~\ref{O7O6model} is a synthetic 1-year time series of ${\rm O}^{7+}/{\rm O}^{6+}$ data (described in further detail in Section~\ref{SS:ModelData}), into which we introduce the observed 2008 data gaps. The filler signal, $F(t)$, shown in red in the middle panel of Figure~\ref{O7O6model}, is a simple linear interpolation across the data gaps. The filler signal form over an $N$-point gap between good data points ${\rm D}_1$ and ${\rm D}_2$ is given by ${\rm F}_n = {\rm D}_1 + ({\rm D}_2 - {\rm D}_1)(n/(N+1))$ for $n = 1$ to $N$. The bottom panels of Figure~\ref{O7O6model} plot the composite time series across two Carrington rotations equivalent to those shown in Figure~\ref{datafig}. 

The wavelet integral transform is linear in the input signals. For a linear superposition of input signals, $T(t) = D(t) + F(t)$, the wavelet transform in a given basis, $\psi \left( t , t' , s \right)$, of the total signal is simply the linear superposition of the wavelet integral transforms of the component signals.
\begin{eqnarray} \label{E:LinearWaveletIntegral}
%\begin{array}
\displaystyle
W_\text{T} \left( t , s \right) & = & \int_{I_{T}} \text{T} \left( t' \right) \psi \left( t , t' , s \right) dt'  \nonumber \\
\displaystyle
W_\text{T} \left( t , s \right) & = & \int_{I_{T}} \left( \text{D} \left( t' \right) + \text{F} \left( t' \right) \right) \psi \left( t , t' , s \right) dt'  \nonumber \\
\displaystyle
W_\text{T} \left( t , s \right) & = & W_\text{D} \left( t , s \right) + W_\text{F} \left( t , s \right) 
%\end{array}
\end{eqnarray}

\noindent Where the integration is taken over the entire set, $I_{T} \subset \mathbb{R}$.

In the most general case, the wavelet transform given by equation (\ref{E:LinearWaveletIntegral}) is a complex number, $W_{T} : I_{T} \times I_{S} \rightarrow \ \mathbb{C}$, where $t \in I_{F} \subset \ \mathbb{R}$ is the full time interval, and $s \in I_{S} \subset \ \mathbb{R}$ is the timescale interval, and is given by,
\begin{equation} \label{E:ComplexWavelet}
W_{T} \left( t , s \right) = \mathfrak{Re} \lbrace W_{T} \left( t , s \right) \rbrace + i\ \mathfrak{Im} \lbrace W_{T} \left( t , s \right) \rbrace
\end{equation}

The (rectified) wavelet power signal for the same time and timescale intervals is the square of the amplitude of the wavelet transform, $\mathcal{P}_{T} : I_{T} \times I_{S} \rightarrow \ \mathbb{R}$.
\begin{equation} \label{E:WaveletPower}
%\begin{split}
\mathcal{P}_{T} \left( t , s \right) = \vert s \vert^{-1} \ \vert W_{T} \left( t , s \right) \vert^2 = \vert s \vert^{-1} \ \left[ \mathfrak{Re}^2 \lbrace W_{T} \left( t , s \right) \rbrace + \mathfrak{Im}^2 \lbrace W_{T} \left( t , s \right) \rbrace \right] \\ 
%\end{split}
\end{equation}

\noindent In general, the real and imaginary components may take on positive, zero, and negative values, and the square of the real and imaginary components ensures the (rectified) total wavelet power spectrum is positive, semi-definite (i.e., non-negative), for all $\left( t , s \right) \in I_{T} \times I_{S}$.

Substituting the signal decomposition of equations (\ref{E:LinearWaveletIntegral}), the amplitude of the wavelet power constructed from a superposition of signals necessarily involves not only the power amplitudes of the individual component signals, but also interference effects between the signals,
\begin{equation} \label{E:NonLinearWaveletPower}
\mathcal{P}_{T} \left( t , s \right) = \mathcal{P}_{D} \left( t , s \right) + \mathcal{P}_{F} \left( t , s \right) + \mathcal{P}_{I} \left( t , s \right)
\end{equation}
\noindent Where we have defined the data signal power, filler signal power, and interference power by,
\begin{equation} \label{E:PowerDecomposition1}
\mathcal{P}_{D} \left( t , s \right)  \equiv \  \vert s \vert^{-1} \ \vert W_{D} \left( t , s \right) \vert^2 = \vert s \vert^{-1} \left[ \mathfrak{Re}^2 \lbrace W_{D} \left( t , s \right) \rbrace + \mathfrak{Im}^2 \lbrace W_{D} \left( t , s \right) \rbrace \right] 
\end{equation}
\begin{equation} \label{E:PowerDecomposition2}
\mathcal{P}_{F} \left( t , s \right)  \equiv \  \vert s \vert^{-1} \ \vert W_{F} \left( t , s \right) \vert^2 = \vert s \vert^{-1} \left[ \mathfrak{Re}^2 \lbrace W_{F} \left( t , s \right) \rbrace + \mathfrak{Im}^2 \lbrace W_{F} \left( t , s \right) \rbrace \right]
\end{equation}
\begin{equation} \label{E:PowerDecomposition3}
\begin{split} 
\mathcal{P}_{I} \left( t , s \right) & \equiv 2 \ \vert s \vert^{-1} \left( \ \mathfrak{Re} \lbrace W_{D} \left( t , s \right) \rbrace \ \mathfrak{Re} \lbrace W_{F} \left( t , s \right) \rbrace \right) \\
& + 2 \ \vert s \vert^{-1} \left( \ \mathfrak{Im} \lbrace W_{D} \left( t , s \right) \rbrace \ \mathfrak{Im} \lbrace W_{F} \left( t , s \right) \rbrace \right)
\end{split}
\end{equation}

\begin{figure}
\epsscale{0.8}
\plotone{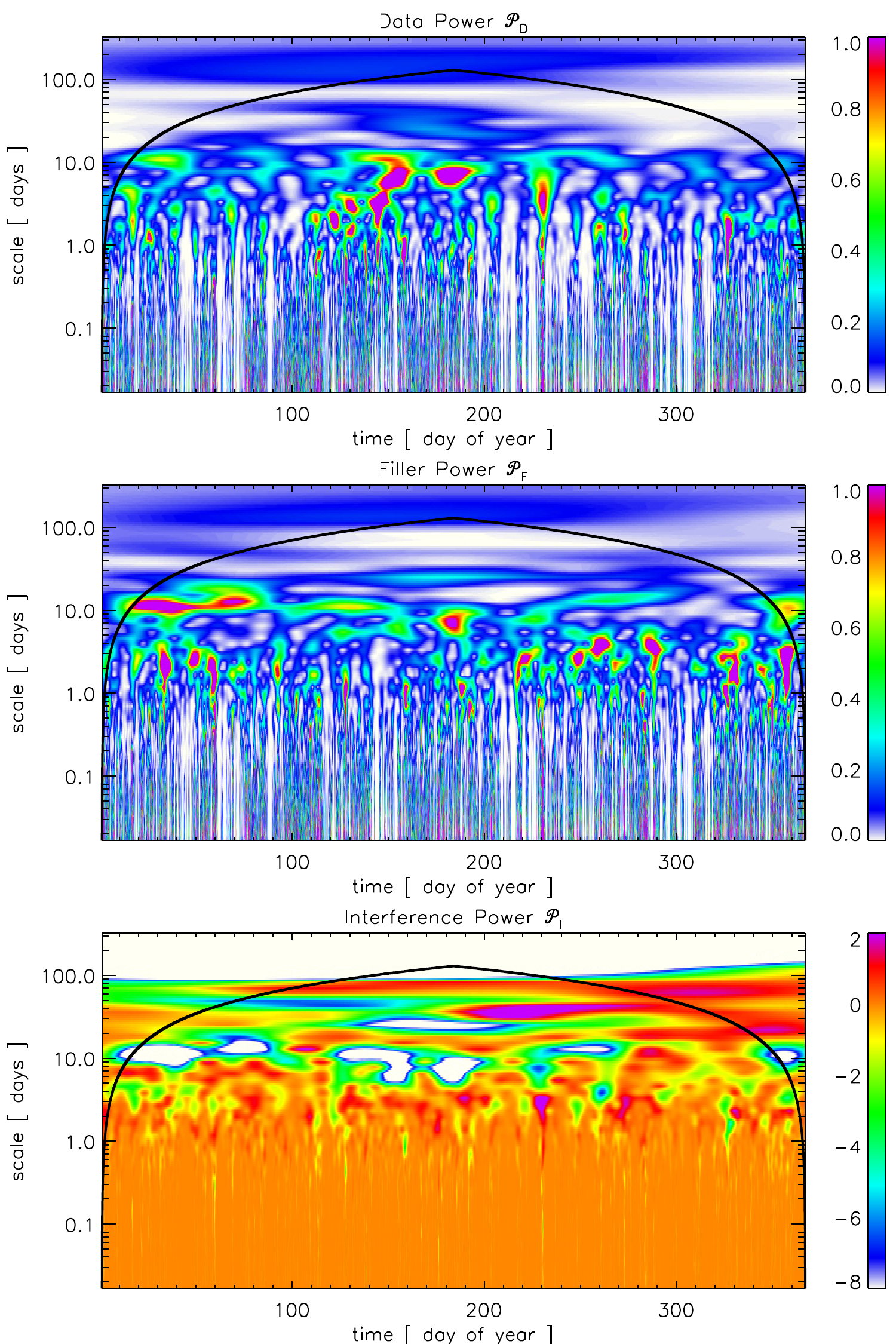}
\caption{The wavelet power spectra for the data signal $\mathcal{P}_{D}$ (top panel), the data gap filler signal $\mathcal{P}_{F}$ (middle panel), and the interference signal $\mathcal{P}_{I}$ (bottom panel).
}
               \label{WaveletPowerDecomposition1}
\end{figure}

Figure~\ref{WaveletPowerDecomposition1} plots the wavelet power spectra decomposition for the time series used in Figure \ref{O7O6model}: for the data signal $\mathcal{P}_{D}$ (top panel), the filler signal $\mathcal{P}_{F}$ (middle panel), and the interference signal $\mathcal{P}_{I}$ (bottom panel).

From equations (\ref{E:WaveletPower}), (\ref{E:PowerDecomposition1}) and (\ref{E:PowerDecomposition2}), the sets of values realized by the total signal power, the data signal power, and the filler signal power spectrograms are all bounded and non-negative, $\mathcal{P}_{T} \left( t , s \right) \geq 0$, $\mathcal{P}_{D} \left( t , s \right) \geq 0$, and $\mathcal{P}_{F} \left( t , s \right) \geq 0$, for all $\left( t , s \right) \in I_{T} \times I_{S}$ (note, the equality holding if and only if the real and imaginary components of the wavelet transform of the particular time series are simultaneously zero). However, for a given $\left( t , s \right) \in I_{T} \times I_{S}$, the real and imaginary components of the respective data and filler transforms may not be of a similar sign, and thus the respective cross terms may be negative. Therefore, in general, the interference power, $\mathcal{P}_{I} \left( t , s \right)$, of equation (\ref{E:PowerDecomposition3}) may realize all real values (positive, zero, and negative).

The negative values of the interference power are interpreted simply as destructive interference, reducing the strictly constructive sum of the individual data and filler signal power spectra such that the total wavelet power spectrum remains a physically meaningful non-negative value. To prove this assertion, we note the decomposition of the time series into measured data and no-measurement filler signals leads to the decomposition of the total wavelet power spectrum given by equation (\ref{E:NonLinearWaveletPower}). By equation (\ref{E:WaveletPower}), the total wavelet power is positive, semi-definite, $\mathcal{P}_{T} \left( t , s \right) \geq 0$, for all $\left( t , s \right) \in I_{T} \times I_{S}$, thus the decomposition of equation (\ref{E:NonLinearWaveletPower}) must be positive, semi-definite for all $\left( t , s \right) \in I_{T} \times I_{S}$,
\begin{equation} \label{E:PositiveSemiDefiniteDecomposition_1}
\mathcal{P}_{D} \left( t , s \right) + \mathcal{P}_{F} \left( t , s \right) + \mathcal{P}_{I} \left( t , s \right) \geq 0 \\
\end{equation}

It is sufficient to show condition (\ref{E:PositiveSemiDefiniteDecomposition_1}) holds for all $\left( t , s \right) \in I_{T} \times I_{S}$. For any fixed $\left( t_{0} , s_{0} \right) \in I_{T} \times I_{S}$, the values realized by the data and filler power spectra are, by equations (\ref{E:PowerDecomposition1}) and (\ref{E:PowerDecomposition2}) respectively, $\mathcal{P}_{D} \left( t_{0} , s_{0} \right) = M \geq 0$ and $\mathcal{P}_{F} \left( t_{0} , s_{0} \right) = N \geq 0$. Additionally, their sum is positive, semi-definite, $\mathcal{P}_{D} \left( t_{0} , s_{0} \right) + \mathcal{P}_{F} \left( t_{0} , s_{0} \right) = M + N \geq 0$ (the equality holds if and only if both $M = 0$ and $N = 0$). In the case $\left( t_{0} , s_{0} \right)$ correspond to a positive or zero interference power value, $\mathcal{P}_{I} \left( t_{0} , s_{0} \right) = P \geq 0$, condition (\ref{E:PositiveSemiDefiniteDecomposition_1}) is trivially satisfied. In the case $\left( t_{0} , s_{0} \right)$ correspond to a negative interference power, $\mathcal{P}_{I} \left( t_{0} , s_{0} \right) = P < 0$, condition (\ref{E:PositiveSemiDefiniteDecomposition_1}) may be written,
\begin{equation} \label{E:PositiveSemiDefiniteDecomposition_2}
\begin{array}{c}
\vert M \vert + \vert N \vert - \vert P \vert \geq 0 \\
\vert M \vert + \vert N \vert \geq \vert P \vert \\
\end{array}
\end{equation}
\noindent Since the choice of fixed $\left( t_{0} , s_{0} \right) \in I_{T} \times I_{S}$ is arbitrary, the assertion is proved for all $\left( t , s \right) \in I_{T} \times I_{S}$.

We note, the power decomposition of equation (\ref{E:NonLinearWaveletPower}) constrains the form of the filler signal power, and subsequently the interference power, to be comparable with that of the data signal power. For a general signal, the wavelet power amplitude distribution at a given timescale depends on the relative magnitude of the range of values over which the signal is distributed. If a particular filler signal model extends the total signal range too far, then the total wavelet power spectrum will be dominated by filler signal and interference effects, completely saturating the measured signal. This constraint requires the range of values of the filler signal model to be at least of similar order as those of the measured signal (examples include, the mean or RMS values of the measured signal, a bounded linear or spline interpolation between measured data points.). In this paper, we compare Linear Interpolation filler signal and a constant Mean Value filler signal.

\subsection{Comparison Power Spectrum and Confidence Levels}

We seek to quantify the new information introduced into the total wavelet power spectrum with the choice of filler signal, by constructing a cofidence level against the null-hypothesis that a given feature in the total wavelet power spectrum is the result of the filler signal and/or nonlinear interference effects. In other words, by filling the no-measurement gaps with a filler signal of arbitrary form we are introducing new information into the system. We aim to quantify the influence of the new information in overall the power spectrum, and thereby elucidate the physical information contained in the (incomplete) measured signal to the highest possible cadence.

\citet{TorrenceCompo1998} discuss stationary significance tests for both red-noise and white-noise by equating a weighted local wavelet power spectrum distribution to an assumed (normal) probability distribution, and then calculating the confidence level according to the particular assumed distribution. \citet{Lachowicz09} offered a prescription to construct a significance level for wavelet power spectra against a time series that is the realization of some physical process that generates a signal with an intrinsic power law, $f^{-\alpha}$, variability. The main underlying assumption is that the Fourier power spectra of the comparison signal approximates that for the given signal. In the case of solar wind composition data, we have no \textit{a priori} reason to suspect that a particular ion (or ions in the case of a composition ratio) are generated by a process with an intrinsic power law variability. Thus, we are not interested in comparing against some physical process governed by (say) an intrinsic red-noise power law, but rather simply looking to quantify the effects of both the (arbitrary) filler signal and its associated interference in the total wavelet power spectrum. Thus, the assumption of a comparison of the standard Fourier power spectra between the two signals is no longer physically relevant.

We construct a statistical confidence level, based on the prescription of \citet{Lachowicz09}, against the null hypothesis that a particular feature in the total power spectrum is due to either the filler signal, a nonlinear interference effect, or a combination of both; equivalently, that a particular feature in the total signal power spectrum is significant as the result of coherent structures in the measured data signal. Thus, we seek a quantitative comparison measure of the structures of the total signal power against the power spectrum consisting of both the filler signal power and interference power. From equation (\ref{E:NonLinearWaveletPower}), we define the comparison power to be,
\begin{equation}
\mathcal{P}_{C} \left( t , s \right) \equiv \mathcal{P}_{F} \left( t , s \right) + \mathcal{P}_{I} \left( t , s \right)
\label{E:ComparisonPower}
\end{equation}

Recall, that while the power of the total power signal is strictly non-negative, $\mathcal{P}_{T} \left( t , s \right) \geq 0$, the range of values of the comparison power spectrum, $\mathcal{P}_{C} \left( t , s \right)$, will, in general, cover some bounded interval that includes zero in the interior, the bounding values of which depend on the relative values of $\mathcal{P}_{F} \left( t , s \right)$ and $\mathcal{P}_{I} \left( t , s \right)$. In other words, the comparison signal includes destructive interference terms of a larger magnitude that the positive filler signal power. That the comparison power may realize negative values requires us to consider the situation in which at a given timescale the comparison power may be so dominated by destructive interference that the resulting confidence level will also realize a negative value. Such an operation is meaningless, since in some sense, it is a comparison between `coherent structures' in the data signal with the process of destructive interference between the data and filler signals.

To rectify this, we use the fact that the comparison power spectrum is bounded from below, $M \equiv \text{inf} \lbrace \mathcal{P}_{C} \left( t , s \right) \rbrace \ \text{for all} \ \left( t , s \right) \in I_{T} \times I_{S}$. We note, in general, $M < 0$, thus, we construct an adjusted total power structure by adding the absolute value of this constant to both sides of equation (\ref{E:NonLinearWaveletPower}).
\begin{equation}
\mathcal{P}_{T} \left( t , s \right) + \vert M \vert = \mathcal{P}_{D} \left( t , s \right) + \mathcal{P}_{C} \left( t , s \right) + \vert M \vert
\label{E:AdjustedPower}
\end{equation} 

Strictly speaking, we are now constructing a confidence level against the null hypothesis that structures in the adjusted total power, $\mathcal{P}_{T} \left( t , s \right) + \vert M \vert$, are the result of structures in the power spectrum of the adjusted comparison signal, $\mathcal{P}_{C} \left( t , s \right) + \vert M \vert$. We ascribe no physical interpretation to the addition of this constant power value across all timescales. It is required to make the confidence level physically consistent across all possible situations; the idea of comparing physical structures with physical structures by ``translating" the process of destructive interference into physically coherent structures. Mathematically, the addition of a constant does not change the relative structure sizes within the power spectra, and thus a significance level constructed on the adjusted spectra retains the physically meaningful information.

For continuous wavelet basis families there is information overlap between timescales (e.g., the basis family is in general not orthonormal), thus we must construct the $p^{th}$ quantile information as a function of timescale. At each fixed timescale, $s_{0} \in I_{S}$, we assume the adjusted comparison power spectrum, $\mathcal{P}_{C} \left( t , s_{0} \right) + \vert M \vert$, is distributed in time as a bounded continuous random variable and construct a probability distribution function, $\rho \left( P ; s_{0} \right)$, from the histogram of the adjusted comparison power values over the full time interval, $t \in I_{T}$; for notational clarity we include the dependence on the given fixed timescale $s_{0}$. There is some ambiguity as to the proper power histogram bin resolution. Under the continuous variable assumption, the bin resolution, $dP$, must be such that all the structures in the adjusted power spectrum are well resolved at the given timescale $s_{0}$. In practice this can be a very small value and therefore computationally expensive. For this study, $dP$ is on the order of 10$^{-5}$. 

Physically, the probability distribution function, $\rho \left( P ; s_{0} \right)$, is a measure of the relative influence of the (adjusted) comparison power in the (adjusted) total power spectrum at the given timescale $s_{0}$. The $p^{th}$ quantile significant power level at each timescale is given by the power value, $X_{p} \left( s_{0} \right)$, such that $\text{Prob}\left( \; \rho \left( P ; s_{0} \right) \leq X_{p} \left( s_{0} \right) \; \right)$. Formally,
\begin{equation}
\text{Prob}\left( \; \rho \left( P ; s_{0} \right) \leq X_{p} \left( s_{0} \right) \; \right) = \int_{0}^{X_{p} \left( s_{0} \right)} \rho \left( P ; s_{0} \right) \ dP
\label{E:PDF}
\end{equation}

\noindent Note, for each fixed timescale, $s_{0} \in I_{S}$, $X_{p} \left( s_{0} \right)$ is a constant. Therefore, at a given fixed timescale, $s_{0} \in I_{S}$, where the adjusted total power is greater than the power level of the $p^{th}$ quantile, $X_{p} \left( s_{0} \right)$, 
\begin{equation}
\mathcal{P}_{T} \left( t , s_{0} \right) + \vert M \vert \geq X_{p} \left( s_{0} \right)
\label{E:SignificantCondition}
\end{equation}

\noindent we can say with $p^{th} \%$ confidence that the particular power structure is not due to the filler signal, nor an interference effect between the data signal and the filler signal. There are often timescales in which condition (\ref{E:SignificantCondition}) is not satisfied, and thus no (adjusted) total power features are significant with respect to the (adjusted) comparison (filler plus interference) power.

For example, an $80\%$ significance level at each timescale, $s \in I_{S}$, is constructed by (numerically) integrating equation (\ref{E:PDF}) until the integral value of  exceeds 0.8. The corresponding upper-integration limit, $X_{p} \left( s \right)$, at which this condition is met is the 80$\%$ significant power level at that timescale. %We note, for computation sake, if the bin resolution, $dP$, is too course such that $X_{p} \left( s \right) = 0$, set $X_{p} \left( s \right) = dP$. 
Condition (\ref{E:SignificantCondition}) then denotes whether the adjusted total power is significant relative to the adjusted comparison power at the coordinates $\left( t , s \right)$.

To illustrate, we choose a fixed timescale, $s_{0} = 2.133$ days, with nice overall variability in the adjusted total power, $\mathcal{P}_{T} \left( t , s_{0} \right)$. Figure~\ref{WaveletPowerDecomposition2} top panel plots the adjusted comparison wavelet power spectrum, $\mathcal{P}_{C} \left( t , s \right) + |M|$, with a horizontal dashed black line demarcating the (fixed) timescale $s_{0} = 2.133$ days. The bottom panel plots the corresponding PDF, $\rho \left( P ; s_{0} \right)$, of the comparison power signal at the (fixed) timescale $s_{0} = 2.133$ days with the 80\%, 90\%, and 95\% significance power levels, $X_{\lbrace 0.8 , 0.9 , 0.95 \rbrace} \left( s_{0} \right)$, demarcated as vertical red lines. Figure~\ref{WaveletPowerDecomposition3} top panel shows the adjusted total power, $\mathcal{P}_{T} \left( t , s \right) + |M|$, with (fixed) timescale $s_{0} = 2.133$ day marker. The bottom panel plots the adjusted total power signal at the $s=2.133$ day timescale with the 80\%, 90\%, and 95\% significance power levels, $X_{\lbrace 0.8 , 0.9 , 0.95 \rbrace} \left( s_{0} \right)$, marked respectively with horizontal red lines, corresponding to the power levels calculated from the adjusted comparison signal PDF. For every adjusted total power value greater than the chosen significance level, we can say with 80\%, respectively, 90\% and 95\%, confidence that the power associated with that feature is not due to filler signal or interference effects.

We note, similar effects are seen in the case of the same synthetic time series with the same introduced data gaps, and a constant Mean Value filler signal form (see Appendix~\ref{S:Appendix1}). Qualitatively, despite the differences between the Linear Interpolation filler signal and constant Mean Value filler signal, the adjusted comparison power spectra share many $0^{th}$-order features (cf. Figure~\ref{WaveletPowerDecomposition2} and Figure~\ref{fA2}). Thus, it is the locations and durations of the data gaps, and therefore the interference power $\mathcal{P}_{I} \left( t , s \right)$ that dominates the (adjusted) comparison power spectra, $\mathcal{P}_{C}  \left( t , s \right) + |M|$; as opposed to the particular filler power spectrum, $\mathcal{P}_{F} \left( t , s \right)$ associated with a particular form of the $F (t)$ signal.

\section{Evaluating Filler Signal Performance with Monte Carlo Ensemble Modeling}
\label{S:MonteCarlo}

We have repeated the procedure described in Section~\ref{S:DataReductionScheme} for an ensemble of 100 different realizations of synthetic ${\rm O}^{7+}/{\rm O}^{6+}$ time series that have the observed 2008 data gaps imposed on each realization. 
In this section we compare results obtained for the Linear Interpolation filler signal (e.g., Figure \ref{O7O6model}) described previously and a constant Mean Value filler signal (e.g., Figure \ref{fA1}) where every missing data point is set to the average value of the measurement data points. 
From these three ensemble sets (the ideal data gap-free model, and the two cases with imposed gaps filled with Linear Interpolation and Mean Value filler forms), we compute the wavelet power spectra for every realization, as well as the 80\% confidence level for both filler signal cases (see Appendix~\ref{S:Appendix1} for representative Figures \ref{fA1}, \ref{fA2}, and \ref{fA3} corresponding to construction of the power spectra confidence levels for the Mean Value filler signal). 
From the individual wavelet power spectra for each of the three ensemble sets, we calculate the mean time-integrated power spectra across all (fixed) timescales, as well as the standard deviation. This ensemble-averaged time-integrated power per scale of the ideal set (the synthetic data without the imposed data gaps) is used to compare with the results of the ensemble-averaged time-integrated power per scale \textit{above the 80\% significance level} computed for each of the synthetic data sets with gaps and their respective filler signal.

Our application of Monte Carlo modeling can be thought of as a mechanism for investigating the particular frequency response or transfer function of some unknown ``black box" in the traditional signal processing sense. The ideal (gap-free) synthetic data corresponds to a set of input waveforms that yield a certain ensemble-averaged, global time-integrated power per scale spectrum. The presence of data gaps, our choice of values to fill those gaps, and our power spectra confidence level threshold condition result in a set of output waveforms which have a well-defined, quantified significance and their own (potentially very different) ensemble-averaged, global time-integrated power per scale spectrum. Understanding and characterizing the influence of missing data on features and properties of the wavelet power spectra is an important and necessary step towards linking those features and properties with the underlying physical processes of their origin.

\subsection{Modeling Synthetic ${\rm O}^{7+}/{\rm O}^{6+}$ Time Series}
\label{SS:ModelData}

In order to use Monte Carlo techniques to evaluate the performance of the different filler signals used to replace missing data, we must have a procedure for generating model time series. Obviously, the model time series should be constructed to have statistical properties as similar to the observations as possible, and in our case here, the ${\rm O}^{7+}/{\rm O}^{6+}$ composition ratio data. Due to its intrinsic variability, a number of authors have suggested that solar wind ionic composition measurements can be reasonably approximated by a first-order Markov process \citep[e.g.,][]{Zurbuchen00,Hefti2000}.
Therefore, we construct a random process with the following recursion
\begin{equation}
Z_n = Z_{n-1}{\rm exp}\left[ -\Delta t/\tau_{1/e} \right] + G_n
\end{equation}
where $\Delta t$ in the exponential decay term is the resolution of the data (12~min) and $G_n$ is a random number drawn from a normalized Guassian distribution. For the \citet{Zurbuchen00} $e$-folding time of $\tau_{1/e} = 0.42$~days ($\sim$10 hours), the exponential decay term describing how much memory the process retains is close to unity for the 12-minute data ($e^{-0.02} \sim 0.98$) and slightly less if we were to model the 2-hour averages ($e^{-0.20} \sim 0.82$). 
The model composition time series is then computed as,
\begin{equation} \label{E:THZModel}
Y_n = {\rm exp}[ \sigma_\ell \hat{Z_n} + \mu_\ell ],
\end{equation}
where $\hat{Z_n}$ is $Z_n$ normalized to unit variance and $\mu_\ell$, $\sigma_\ell$ are the mean and standard deviation of the natural logarithm of the measured ionic composition ratio. \citet{Zurbuchen00} showed that the ${\rm O}^{7+}/{\rm O}^{6+}$ data had a log-normal distribution with ($\mu_\ell$, $\sigma_\ell$)=($-$1.32, 0.45) and we use those values here. Our model time series reproduces the 10-hour $e$-folding time of the autocorrelation function and has an FFT power spectra that falls off between $f^{-1}$ and $f^{-2}$, consistent with the \citet{Zurbuchen00} analysis. 

We note that in the \citet{Edmondson2013} companion paper, we present the results of this modeling tuned to the ${\rm C}^{6+}/{\rm C}^{4+}$ ionic charge state ratio. There we show that, while this type of Markov process modeling produces the log-normal distribution of the in-situ measurements (by construction), the global time-integrated power per scale spectra of the observations contains real information about the physical structure and dynamics of their source region, including properties of the plasma and coronal magnetic field, that are not and cannot be accounted for by a purely random process.

\subsection{Ensemble-Averaged Integrated Power per Scale}
\label{SS:AvgIPPS}

\begin{figure}
\epsscale{0.75}
\plotone{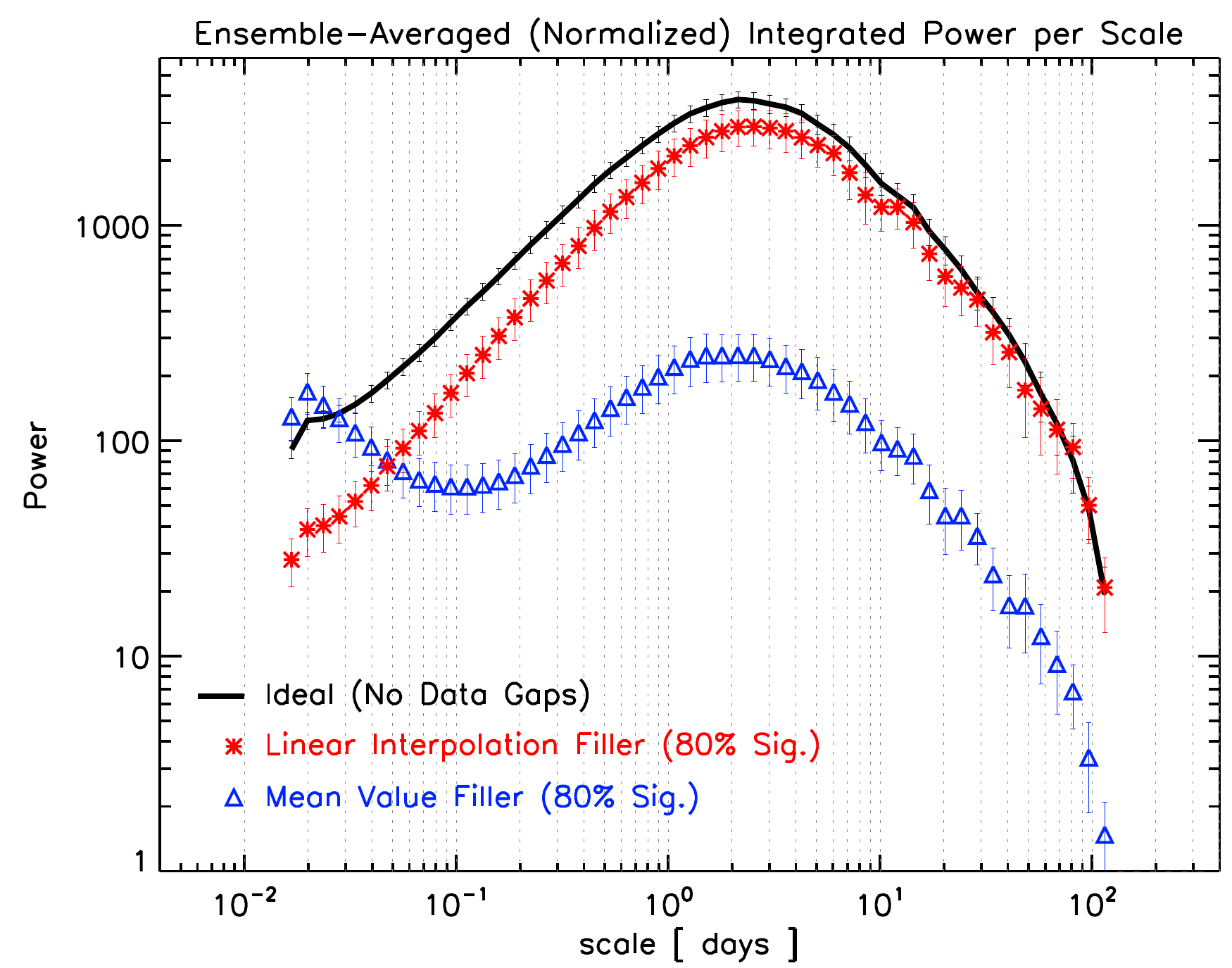}
\caption{Comparison of the Monte Carlo ensemble-averaged, normalized, time-integrated power per scale spectra. The black solid line shows the ideal, data gap-free case. The Linear Interpolation filler results are shown as red asterisks and the constant Mean Value filler results are shown as blue triangles. Error bars indicate the statistical variation in the Monte Carlo ensemble set in each timescale bin. Each of the data plus filler signal curves represent the integrated power above the 80\% significance level derived from the filler and interference power calculated in each model realization.
}
               \label{FillerPerformance}
\end{figure}

Figure~\ref{FillerPerformance} plots the Monte Carlo simulation results for our three ensemble set averages. The ideal case (with no missing data) is shown as the black line; all of the wavelet power of each realization is deemed significant because there are no data gaps. Thus, the ideal integrated power per scale represents the ensemble-average of the global periodicities (Fourier modes) of the `input waveforms'. The integrated power per scale for the ensemble-average Linear Interpolation (red asterisks) and ensemble-average Mean Value (blue triangles) cases are the `output waveforms' that result from taking the ideal set of Monte Carlo realizations, adding the 2008 data gaps and a particular filler signal, and applying the 80\% power spectra confidence level threshold condition. In other words, the ensemble-average global periodicities (Fourier modes) above the 80\% significance level. The error bars in each color represent the statistical uncertainty of one standard deviation in each timescale bin for each of the ensemble sets.   

We see that for the Linear Interpolation case, the shape of the ideal cases' integrated power per scale is well preserved for $s \gtrsim 1$~day but shows increasing departure from the the ideal case at increasingly smaller time scales. The overall relative shape of the global periodicities are qualitatively similar, but the Linear Interpolation case is increasingly attenuating the `input waveform' power for $s < 1$~day. The Mean Value filler ensemble-averaged results however show a very different spectrum response. While there is much more attenuation across all scales, the behavior for scales $s \gtrsim 1$~day retains the shape of both the ideal and Linear Interpolation cases. However, for scales below $\sim$0.1~days (2.4~hrs), the Mean Value ensemble-averaged integrated power per scale rebounds from a local minimum and increases in magnitude through to the Nyquist frequency of the time series. 

The Mean Value filler case's spectral response for $s \lesssim 0.1$~days (2.4~hrs) is primarily the signature of the occurrence frequency of the data gap durations. This small scale (high frequency) amplification is a direct spurious effect following from the interleaving of the constant filler signal within gap durations with relatively high frequency of occurrence, and the measured data signal. The smallest duration data gaps occur ubiquitously throughout the measured data signal, and filling these gaps with any constant value has the effect of creating spurious small scale, pulse-like structures in the full (data plus filler) signal as the wavelet transform passes through the small scales. At wavelet transform scales larger than the majority of gap durations, this effect is mitigated as the gap durations become much smaller than the wavelet filter band pass, hence the integrated power per scale shape reflects that of the ideal gap-free case. In this example, the distribution of 2008 gap durations in the 12 minute ${\rm O}^{7+}/{\rm O}^{6+}$ measurements (Figure~\ref{MissingDataPDF}) indicates the largest gap is $\sim$2.5 days, the time scale above which the integrated power per scale for all three cases reflect similar trends. Additionally, the vast majority of gap durations, $90.4$\%, occur at timescales below 0.1~days in duration, at which point the spurious high-frequency effect dominates. Finally, there is a transition zone between $\sim$0.1 and $\sim$2.5 days in which the slope is much shallower than the ideal gap-free case.

On the other hand, with the Linear Interpolation filler signal, we are able to maintain the relative shape of the ideal data gap-free ensemble-averaged integrated power per scale spectrum over a broader range of scales but with increasing attenuation at progressively smaller scales $s \lesssim 1$~day. This is essentially due to the fact that, in any given data gap, the difference between filler signal values and neighboring synthetic model values are, by construction, much closer (as opposed to the synthetic model values and Mean Value filler). 

\begin{figure}
\epsscale{0.75}
\plotone{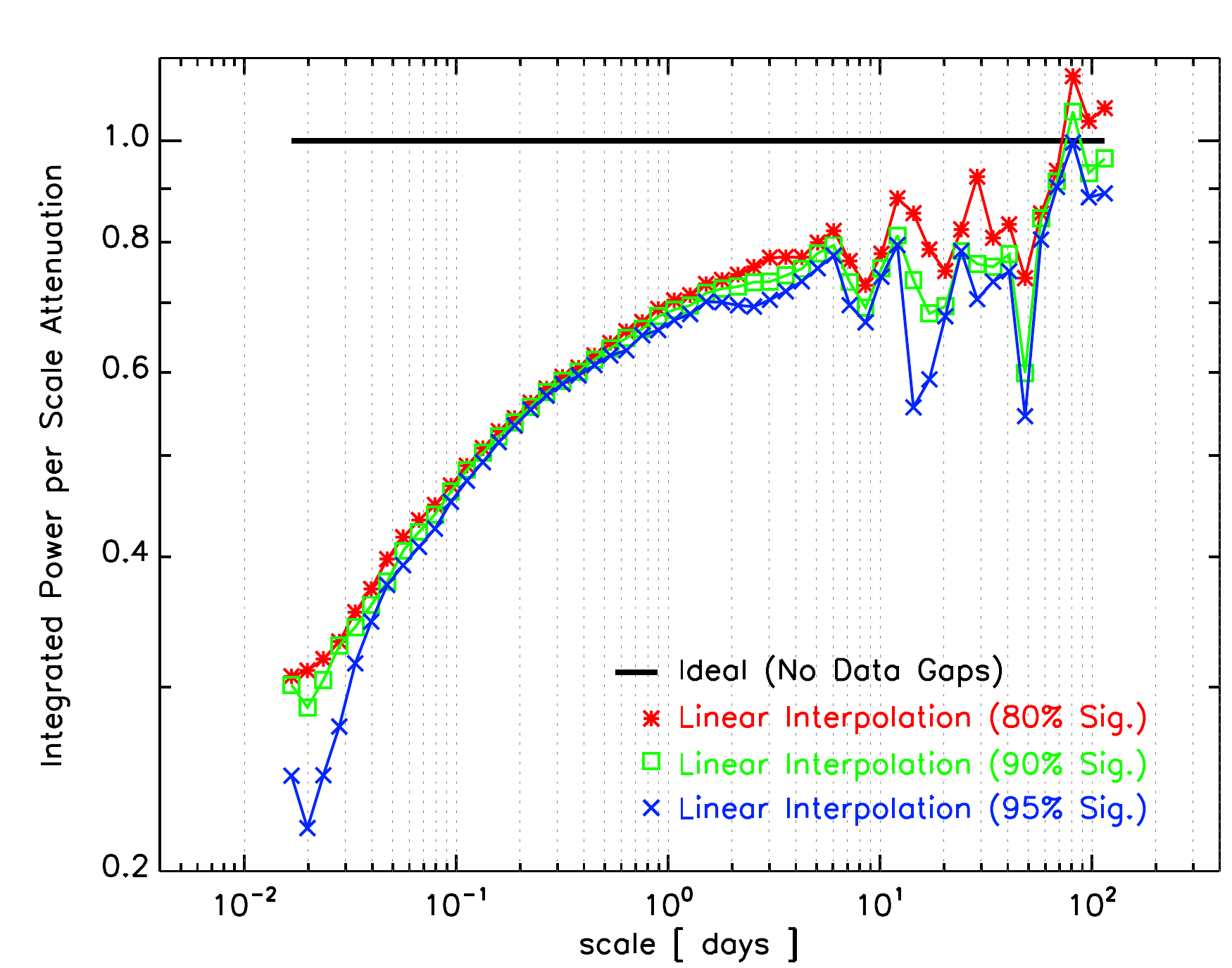}
\caption{Evaluation of the ensemble-averaged integrated power per scale attenuation from the ideal data gap-free case. Here, three significance levels against the comparison power for the Linear Interpolation filler signal are used:  80\% (red), 90\% (green), and 95\% (blue). 
%The overall trend of more attenuation with a higher significance level threshold is relatively minor compared to the dominant feature of increased attenuation at spatial scales associated with the most frequent missing data duration.
}
               \label{HybridFillerPerformance}
\end{figure}

Figures~\ref{WaveletPowerDecomposition2} and \ref{WaveletPowerDecomposition3}, illustrate the construction of significance levels at 80\%, 90\%, and 95\% comparison power. Using this procedure, we calculated the ensemble-averaged integrated power per scale for the Linear Interpolation filler at the 80\% (shown as red asterisks in Figure \ref{FillerPerformance}), as well as the 90\% and 95\% significance levels to examine the attenuation due to the significance level threshold conditions. Figure~\ref{HybridFillerPerformance} plots these results normalized to the ideal gap-free average integrated power per scale spectrum. The ideal case is shown as the black line at unity and the 80\%, 90\%, and 95\% Linear Interpolation cases are shown as red asterisks, green squares, and blue crosses, respectively. Here the scale-dependence of the attenuation with respect to the ideal gap-free ensemble spectrum is readily visible showing a drop from roughly 0.70--0.80 of the ideal average power for $s > 1$~days down to $\sim$0.30 of the ideal power at $s \sim 0.02$~days. For our particular set of model time series and data gap structure, the ensemble-averaged power per scale curves for the different significance levels show very little separation with respect to each other. This could be expected from examination of Figure~\ref{WaveletPowerDecomposition3} where the time-integrated power for the $s=2.133$~day cut shows only minor differences in the total area under the $\mathcal{P}_T+|M|$ curve and above the various significance level thresholds. Therefore, our selection of the 80\% significance level appears reasonable, at least for the time series and data gap properties analyzed here. Mean Value filler comparisons across 80\%, 90\%, and 95\% significance levels exhibit similar trends, albeit with stronger relative attenuation.

\section{Wavelet Analysis of ${\rm O}^{7+}/{\rm O}^{6+}$ 12-Min Data}
\label{S:WaveletO7O6}

\begin{figure}
\epsscale{0.8}
\plotone{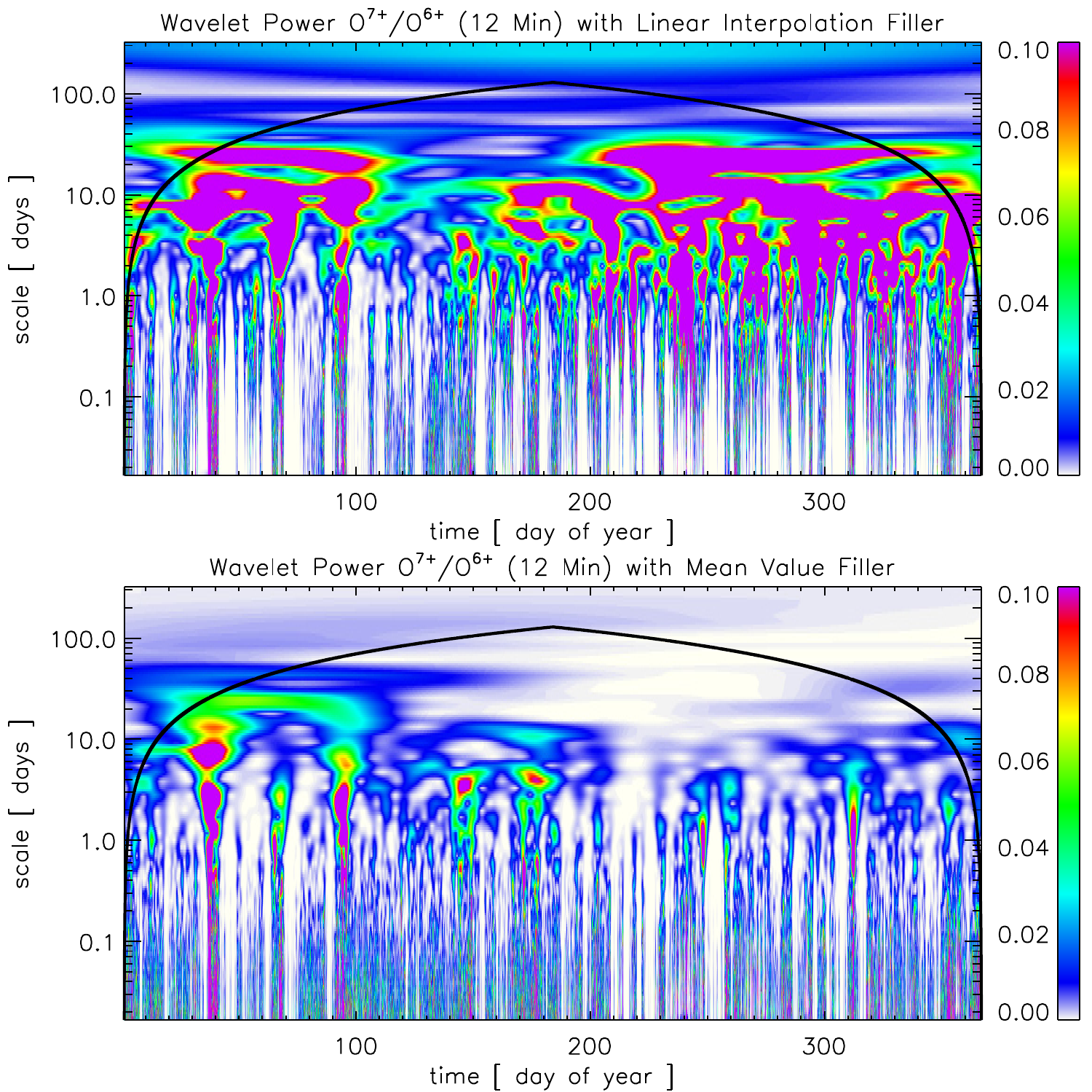}
\caption{Comparison of ${\rm O}^{7+}/{\rm O}^{6+}$ the total wavelet power spectra for Linear Interpolation filler signal (top panel) and constant Mean Value filler signal (bottom panel).
}
               \label{figO7O6wavelet}
\end{figure}

\begin{figure}
\epsscale{0.8}
\plotone{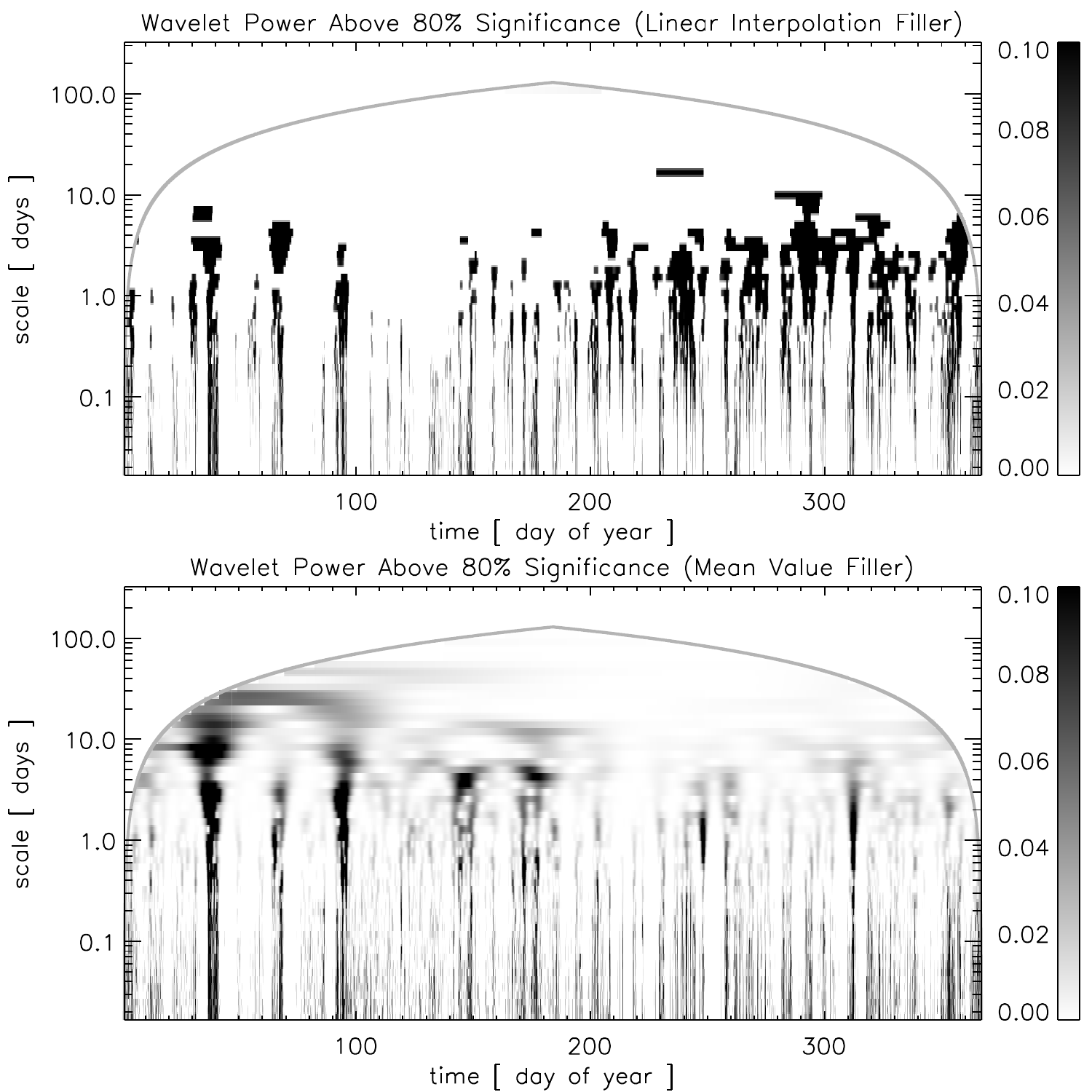}
\caption{The total ${\rm O}^{7+}/{\rm O}^{6+}$ wavelet power $\ge$80\% significance level: The top panel shows the data with the Linear Interpolation filler signal and the bottom panel shows the data with the constant Mean Value filler signal.
}
               \label{figO7O6wavelet2}
\end{figure}

\begin{figure}
\epsscale{1.0}
\plotone{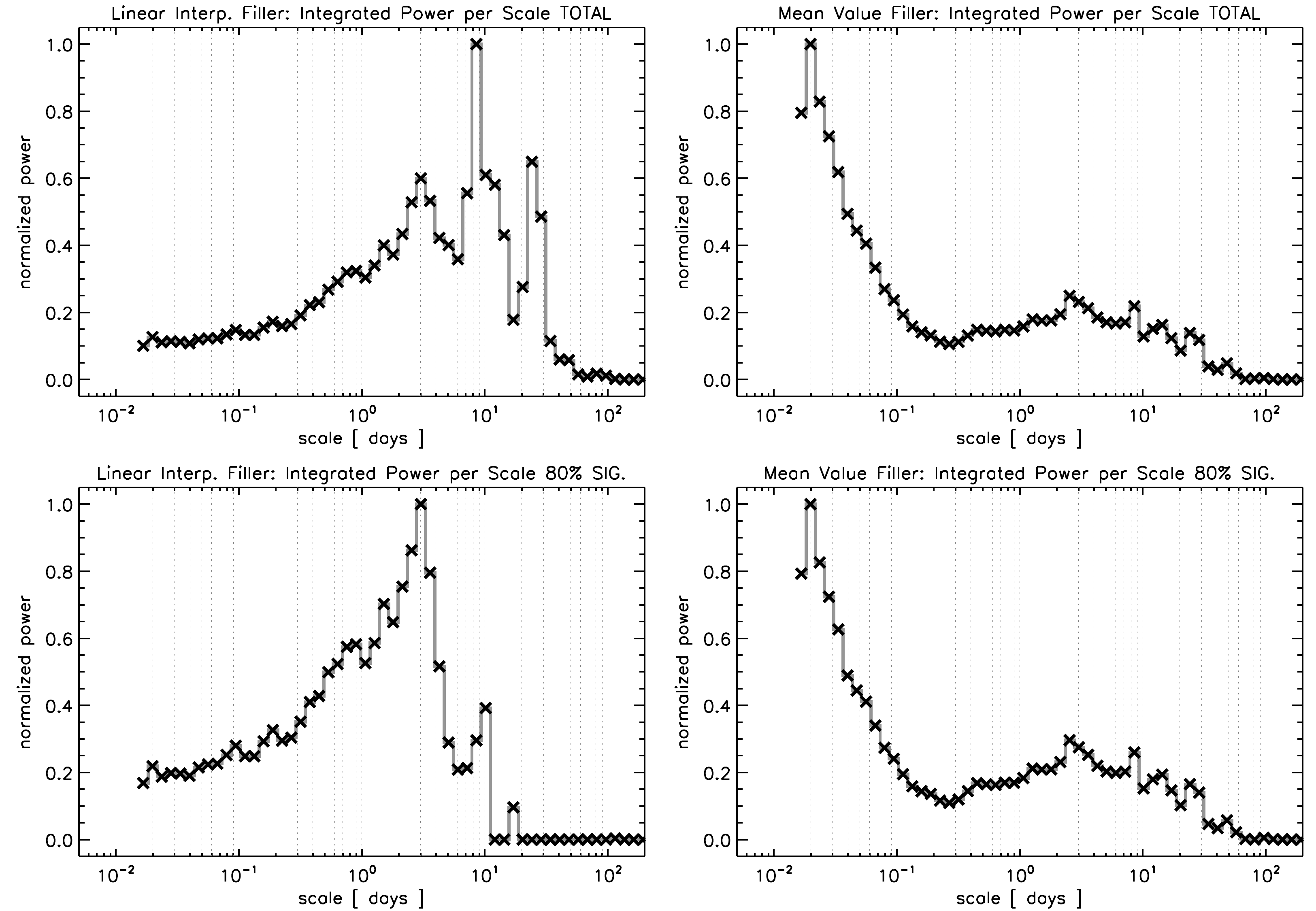}
\caption{Comparison of normalized integrated power per scale of the ${\rm O}^{7+}/{\rm O}^{6+}$ data and the Linear Interpolation filler (left column) or Mean Value filler (right column). Top row plots the \textit{total} integrated power per scale for each filler signal case (from Figure~\ref{figO7O6wavelet}) and the bottom row plots the integrated power \textit{above the 80\% significance level} per scale (from Figure~\ref{figO7O6wavelet2}).
}
               \label{figO7O6wavelet3}
\end{figure}

We have applied the analysis procedure outlined in Section~\ref{S:DataReductionScheme} to the ACE/SWICS ${\rm O}^{7+}/{\rm O}^{6+}$ data shown in Figure~\ref{datafig} using both the Linear Interpolation and constant Mean Value prescriptions for filling the data gaps. The wavelet power spectra for full data plus both filler signal models were calculated, as well as the wavelet power spectra decomposition from the two filler signals and their respective nonlinear interference components. The adjusted comparison power was then used to construct the 80\% significance levels for each timescale. The results are shown in Figures \ref{figO7O6wavelet}, \ref{figO7O6wavelet2}, and \ref{figO7O6wavelet3}. In Figure \ref{figO7O6wavelet}, the total wavelet power spectra for the ${\rm O}^{7+}/{\rm O}^{6+}$ data with a Linear Interpolation filler and the constant Mean Value filler are shown in the top and bottom panels, respectively.  Figure \ref{figO7O6wavelet2} plots the corresponding wavelet power that exceeds the 80\% confidence level thresholds. Figure \ref{figO7O6wavelet3} plots on a linear scale, the normalized time-integrated power per scale for both the overall total power spectra (top row) and 80\% significant power (bottom row), for the Linear Interpolation filler signal (right column) and constant Mean Value filler signal (left column).

For the Morlet wavelet family, the time-integrated power per timescale (also known as the global wavelet power) of Figure \ref{figO7O6wavelet3} is akin to Fourier mode decomposition. The integrated wavelet power per scales of the ${\rm O}^{7+}/{\rm O}^{6+}$ for both filler signal forms, in both the total and significant power, exhibit a number of well defined peaks corresponding to relatively well defined Fourier modes (globally periodic timescales) in similar timescale neighborhoods.

In the Linear Interpolation filler signal case (left column), there are three strong Fourier modes (peaks) occurring at approximately $\lbrace$ 3, 8--10, 18--28 $\rbrace$~days, in both total and significant power cases. Below $\sim$1~day timescales, it becomes difficult to discern Fourier modes (peaks) from the power law shape. As explained above, the large high-frequency effect (timescales $\lesssim$ 0.1 days) in the Mean Value case (right column) reflects the nature of the gap durations. Outside of this effect, there are well-defined Fourier modes in both the total power and significant power that occur at approximately, $\lbrace$ 2.5, 8--9, 13--17, 25--30, 45--50 $\rbrace$~day timescales.  

The smallest identifiable Fourier modes, at 2.5 and 3 days, respectively, may be an artifact of the largest data gap duration in the measured signal. However, \citet{Zhao09} showed that the slow solar wind, as determined by ${\rm O}^{7+}/{\rm O}^{6+} \ge 0.145$, has a mean width centered on the heliospheric current sheet of approximately 20$^\circ$ (40$^\circ$) during  solar minimum (maximum). We note that our other significant integrated power peak at the  3--4 day correlation timescale corresponds to an $\sim$45$^\circ$ width given the 13$^\circ$~day$^{-1}$ solar rotation rate. This may reflect crossing the slow solar wind region surrounding the helmet streamer belt in a highly inclined configuration and would be consistent with the width of the slow solar wind distribution observed in the Ulysses fast latitude scan \citep[][]{McComas00}. 

\citet{Temmer07} identified 9-day periodicities in ACE solar wind parameters over the 1998-2006 period and showed these likely arose due to the distribution of coronal holes via time series of coronal hole area. \citet{Katsavrias12} likewise identified both the 9-day and 13.5~day peaks in solar wind speed, proton temperature, density, and components of the magnetic field over a four solar cycle interval (1966--2010). The 18--28 day, and 25--30 day timescales are clearly associated with Carrington rotation effects.

Interestingly, the Carrington rotation periodicity is absent from the significant Linear Interpolation filler power spectra. This is largely due to the unusual solar minimum conditions during 2008. In the 12-minute ${\rm O}^{7+}/{\rm O}^{6+}$ data shown in Figure~\ref{datafig}, one may identify a \textit{qualitative} recurrent $\sim$5~day enhancement repeating with a 27-day periodicity for three consecutive Carrington Rotations at the beginning of the year (CR2066--CR2068). However, this enhancement is absent (potentially due to a data gap) in the fourth Carrington Rotation (CR2069) and virtually indistinguishable during the remainder of the year. In fact, the wavelet power in the top panel of Figure~\ref{figO7O6wavelet} also shows this as a reasonably strong intensity stripe at the 27-day timescale that falls outside of the 80\% significance contours for the first 100~days and a more modest intensity signal at that timescale up until 200~days; thus, the signal is present in the total wavelet power, but does not exceed the 80\% confidence level threshold derived from the Linear Interpolation filler signal and its interference effects.

On the other hand, while the 27-day periodicity is absent, its first harmonic at approximately 13.5~days is present above the 80\% significance level for the entire data set duration. The preference for this periodicity seems likely due to the large-scale coronal magnetic field structure and consequently, the solar wind structure in the heliosphere. During the solar cycle 23 solar minimum, the polar fields were substantially weaker than usual resulting in a more highly warped helmet streamer belt, more pseudostreamers, and more complexity in the mapping of the solar wind source regions to smaller, low latitude coronal holes \citep[e.g.,][]{Lee09,Riley12}. \citet{MursulaZieger1996} have argued the 13.5~day periodicity could be explained by a two slow-fast stream structure per Carrington Rotation that may result from a highly warped helmet streamer belt, but it is not clear this is universally applicable \citep[e.g., see discussion by][]{Temmer07}.

Finally, the 45--50 day Fourier mode that shows up in the large timescale tail of the Mean Value filler case, is likely a harmonic of the Carrington rotation periodicity. Above this scale, for a single year data set (2008), the total integrated power per scale is primarily bounded by the wavelet cone-of-influence (see Figure \ref{figO7O6wavelet}).

\section{Conclusions}
\label{S:Conclusions}

We have presented a generalized procedure for constructing a wavelet power spectrum significance level measure that quantifies the relative influence of two interleaved signals. In this investigation, the total signal is an interleaved combination of measured data and some (general, arbitrary) signal imposed to fill the `no measurement' gaps at the given cadence. We constructed a statistical confidence level against the null-hypothesis that a given feature in the total wavelet power spectrum is strictly due to filler signal and the nonlinear interference effects between the filler signal and measured data signal. 

We apply this power spectra confidence procedure on Monte Carlo simulations of synthetic ${\rm O}^{7+}/{\rm O}^{6+}$ ionic composition data to evaluate the performance of the Mean Value and Linear Interpolation filler signals. Using the performance criteria of reproducing the ideal, data gap-free ensemble-averaged time-integrated power per scale, we show that the Linear Interpolation filler signal does a better job than the Mean Value signal across all but the smallest temporal scales and effectively acts as a low-pass filter suppressing the inherent high frequency (small timescale) power that arise from the frequency of the missing data and duration of the data gaps. We show that for our sparsely populated data set, the 80\%, 90\%, and 95\% confidence levels yield almost identical results for the synthetic data ensemble. 

We calculated the ${\rm O}^{7+}/{\rm O}^{6+}$ wavelet power during the quiet-sun solar minimum of 2008 using both the Mean Value and Linear Interpolation filler signals, show the structure of their derived 80\% power confidence levels, and present the total and $\ge$80\% significant time-integrated power per scale spectra. Our analysis using the Linear Interpolation data gap filler signal yields strong Fourier mode harmonics in both the total and significant integrated power per scale spectrum at $\lbrace$ 3, 8--10, 18--28 $\rbrace$ days. Each of these peaks are also visible in the total and significant integrated power per scale when using the Mean Value filler, but the relative magnitude of the spectrum for scales $\gtrsim$0.10~days is dwarfed by the (spurious) power associated with very-high frequencies ($s < 0.10$~days). In a companion publication \citep{Edmondson2013}, we have applied the power spectra confidence analysis presented here to the ${\rm C}^{6+}/{\rm C}^{4+}$ ionic composition ratio and discuss the implications of the coherent structure and variability of ionic composition ratios for current theories of solar wind generation.

\acknowledgments

The authors would like to thank the reviewer for their comments and suggestions which substantially improved the paper.
J.K.E., S.T.L., and T.H.Z. acknowledge support from NASA LWS NNX10AQ61G and NNX07AB99G. B.J.L acknowledges support from AFOSR YIP FA9550-11-1-0048 and NASA HTP NNX11AJ65G.

% Appendix %%%%%%%%%%%%%%%%%%%%%%%%%%
\appendix

\section{Power Spectra and Confidence Levels for the Mean Value Filler Signal}
\label{S:Appendix1}

In Section~\ref{S:DataReductionScheme} we used one realization of the Zurbuchen et al. inspired Markov process modeling to illustrate the wavelet power spectra confidence procedure. First, we generated a synthetic time series time series from Equations (19) and (20), then imposed data gaps corresponding to the missing data intervals in the 2008 12~minute ${\rm O}^{7+}/{\rm O}^{6+}$ data. The analysis of Section~\ref{S:DataReductionScheme} used the Linear Interpolation filler signal to populate the missing data intervals and calculate the various wavelet power spectra associated with the `good measurement' data, the filler signal, and their nonlinear interference. A reference comparison power was constructed from the filler signal and interference power contributions and used to quantify the significance levels associated with features in the total wavelet power spectra. In section~\ref{S:MonteCarlo}, we presented the ensemble-averaged results from performing this procedure on a set of 100 realizations of the Markov process using both the Linear Interpolation filler signal and a constant Mean Value filler signal. 

\begin{figure}
\epsscale{1.0}
\plotone{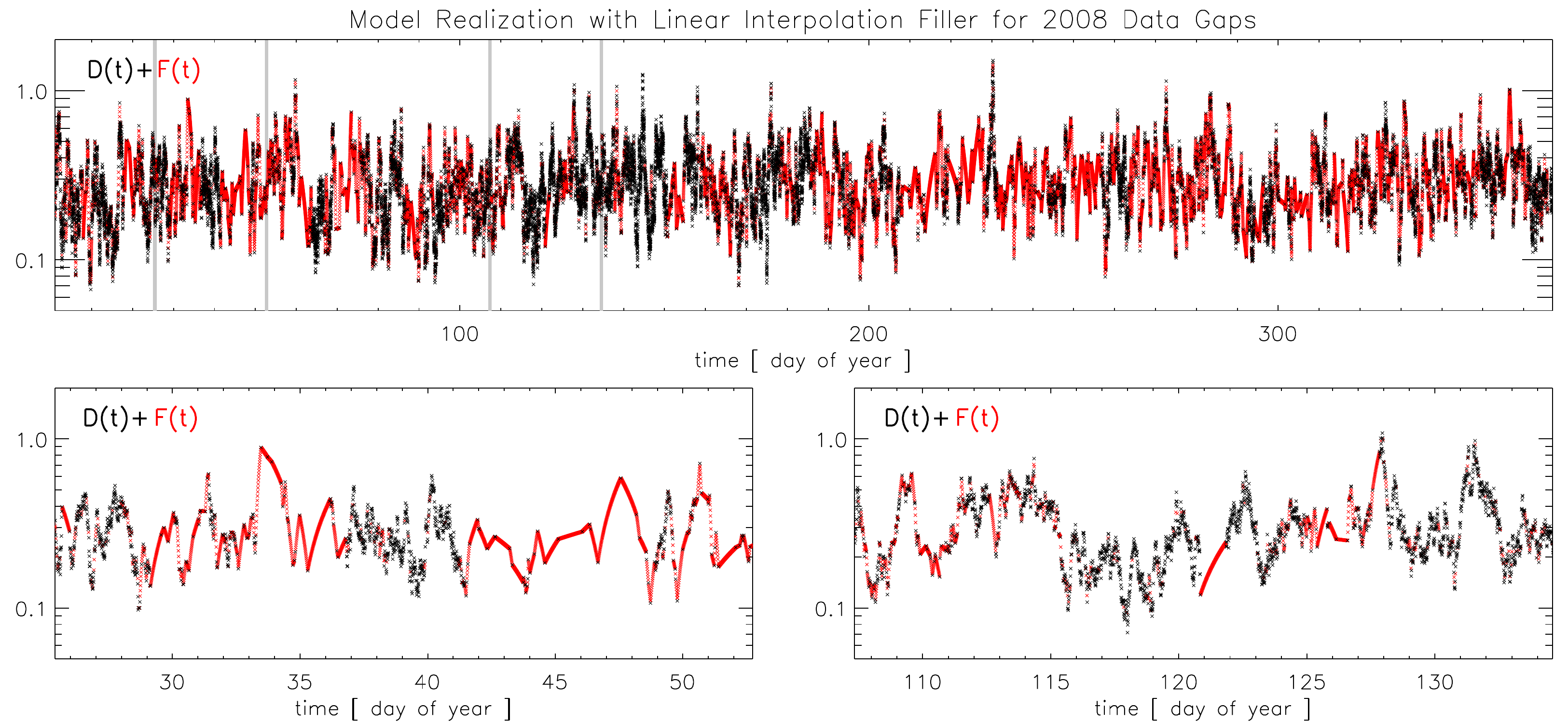}
\caption{Top row, one realization of the \citet{Zurbuchen00} model time series $D(t)$ (plotted in black) with the observed 2008 data gaps imposed and then filled with the Linear Interpolation filler signal $F(t)$ (plotted in red). Bottom row, close-up views of the model data and filler signals during the Carrington Rotation periods shown in Figure~\ref{datafig}.
}
               \label{O7O6model}
\end{figure}

\begin{figure}
\epsscale{0.8}
\plotone{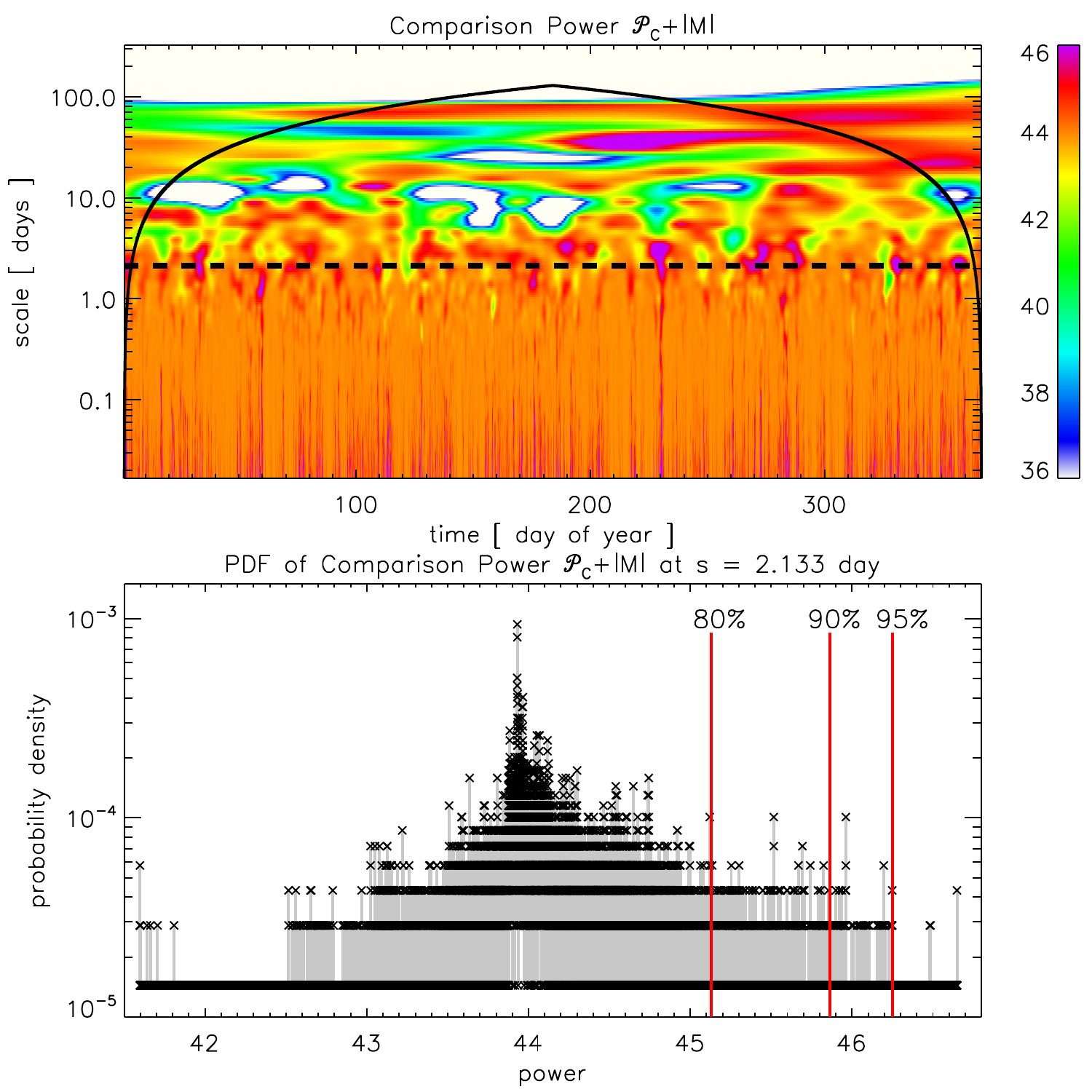}
\caption{The top panel plots the adjusted comparison power spectrum, $\mathcal{P}_{C} + |M|$ (note, the adjusted comparison power spectrum is dominated by the interference power and offset, and thus looks similar to the bottom panel of Figure \ref{WaveletPowerDecomposition1}). The bottom panel shows the probability density function of the comparison signal power at timescale $s=2.133$~day. The 80\%, 90\%, and 95\% significance values are calculated from this PDF and plotted as the vertical red lines.
}
               \label{WaveletPowerDecomposition2}
\end{figure}

Here we present details of our power spectra confidence level construction for the model realization example of section~\ref{S:DataReductionScheme} with the Mean Value filler signal. 
Figure~\ref{fA1} shows time series in two zoomed in views for same synthetic data $D(t)$ shown in Figure~\ref{O7O6model} but for the constant Mean Value filler signal $F(t)$.  
Following the decomposition of the time series and calculation the power of the wavelet transforms of the constituent components, the top panel of Figure~\ref{fA2} plots the comparison power plus offset $\mathcal{P}_{C} + |M|$. The bottom panel of Figure~\ref{fA2} plots the PDF of the comparison power at the $s = 2.133$~day scale and the 80\%, 90\%, and 95\% levels of the distribution. The similarities and differences in the comparison power between the Mean Value filler signal and the Linear Interpolation case are readily apparent when comparing Figures~\ref{fA2} and \ref{WaveletPowerDecomposition2}. 

First, we see that the comparison power wavelet has both a similar range in magnitude and qualitative large scale structure in $(t, s)$. For example, the regions of relatively low comparison power levels (saturated as white in the color scale) at the $s\sim10$~day scale features are quite similar in shape and location, the overall trend of comparison power levels at scales $0.1 \lesssim s \lesssim 1.0$~days being elevated with respect to $s \gtrsim 1.0$~days, and the largest power levels (saturated with magenta in the color scale) corresponding to many fine-scale linear striations for $s \lesssim 0.1$~days. From the relative amount of color scale saturation at the smallest scales, the Mean Value comparison power has a broader temporal extent indicating more high frequency interference power throughout the time series.
The lower panels of Figures~\ref{fA2} and \ref{WaveletPowerDecomposition2} however, have a very different PDFs for their respective comparison power levels (although the overall range of values are comparable). While the Linear Interpolation PDF is symmetric and centered around a mid-point value of $\sim$44, the Mean Value PDF has more of an exponential fall off from a maximum at $\sim$41.3. Thus, we can see the relative contributions to the comparison power from: (1) the form and values of the $F(t)$ filler signal in shape of the PDF at a given scale, and (2) in the location, duration, and distribution of the data gaps (i.e., where $F(t) \ne 0$) and the resulting interference power in the overall, large scale properties of the comparison wavelet power.

\begin{figure}
\epsscale{0.8}
\plotone{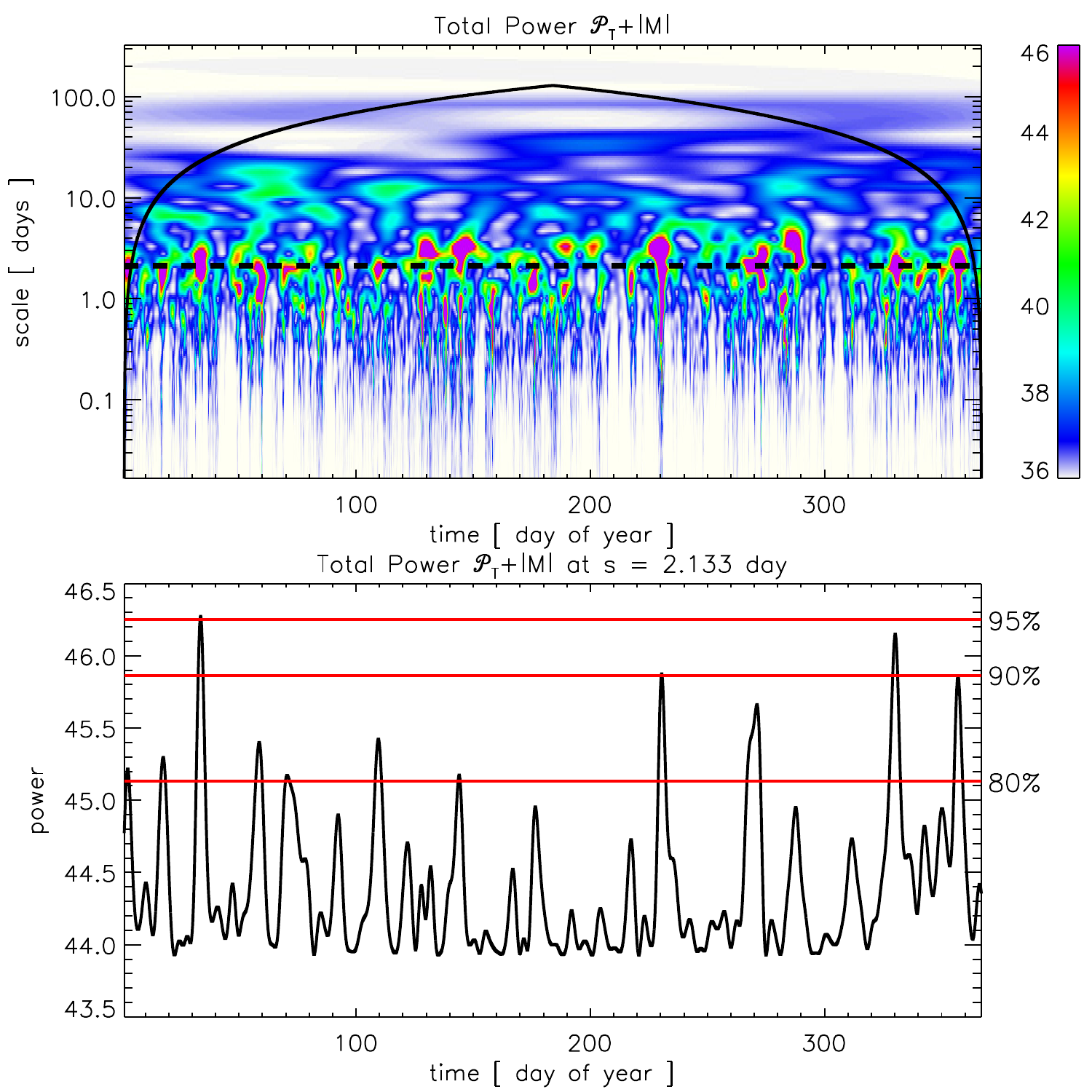}
\caption{Top panel plots the adjusted total power spectrum, $\mathcal{P}_{T} + |M|$. The bottom panel plots the adjusted total power vs. time at fixed $s=2.133$ day with the 80\%, 90\%, and 95\% significance levels over-plotted.
}
               \label{WaveletPowerDecomposition3}
\end{figure}

Figure~\ref{fA3} plots the total power plus offset $\mathcal{P}_{T} + |M|$ in the top panel and the $s = 2.133$~day cut in the lower panel for the Mean Value filler in the same format as and for direct comparison with Linear Interpolation results in Figure~\ref{WaveletPowerDecomposition3}. Again, there are both important similarities in the overall qualitative properties of the wavelet power and important differences arising from the different filler signals. In the wavelet power, the $s \gtrsim 5.0$~days features are less prominent in the Mean Value case, whereas the smallest scale structures at $s \lesssim 0.1$~days are much more prominent. The lower panel of Figures~\ref{fA3} shows that the Mean Value case has no total power at the $s=2.133$~day scale that falls above the 95\% significance level, and only two temporal locations that exceed the 80\% level. The most prominent Mean Value total power feature at $t\sim$DOY~145 is obviously present in the Linear Interpolation wavelet power, but only slightly exceeds the Linear Interpolation 80\% significance level at this scale.  

The overall, qualitative scale-dependent influence of the Mean Value and Linear Interpolation filler signals to the total wavelet power in the single representative synthetic data example from Section~\ref{S:DataReductionScheme} are reproduced in the properties of the ensemble-averaged behavior obtained from the Monte Carlo simulations in Section~\ref{S:MonteCarlo} and in the application to the actual ${\rm O}^{7+}/{\rm O}^{6+}$ measurements presented in Section~\ref{S:WaveletO7O6}.

%% References %%%%%%%%%%%%%%%%%%%%%%%%%%

\clearpage

%% Figures %%%%%%%%%%%%%%%%%%%%%%%%%%

%%%%%%%%%%%%%%%%%%%%%%%%%%%%%%%%%%%%%%%%%%%%%%%%%%%%%%%%	O7/O6 Data
%%%%%%%%%%%%%%%%%%%%%%%%%%%%%%%%%%%%%%%%%%%%%%%%%%%%%%%%
%%%%%%%%%%%%%%%%%%%%%%%%%%%%%%%%%%%%%%%%%%%%%%%%%%%%%%%%	O7/O6 Data Gaps
%%%%%%%%%%%%%%%%%%%%%%%%%%%%%%%%%%%%%%%%%%%%%%%%%%%%%%%%	O7/O6 Data Gaps
%%%%%%%%%%%%%%%%%%%%%%%%%%%%%%%%%%%%%%%%%%%%%%%%%%%%%%%%	Data Model
%%%%%%%%%%%%%%%%%%%%%%%%%%%%%%%%%%%%%%%%%%%%%%%%%%%%%%%%
%%%%%%%%%%%%%%%%%%%%%%%%%%%%%%%%%%%%%%%%%%%%%%%%%%%%%%%%	Data Model
%%%%%%%%%%%%%%%%%%%%%%%%%%%%%%%%%%%%%%%%%%%%%%%%%%%%%%%%
%%%%%%%%%%%%%%%%%%%%%%%%%%%%%%%%%%%%%%%%%%%%%%%%%%%%%%%%	Data Model
%%%%%%%%%%%%%%%%%%%%%%%%%%%%%%%%%%%%%%%%%%%%%%%%%%%%%%%%
%%%%%%%%%%%%%%%%%%%%%%%%%%%%%%%%%%%%%%%%%%%%%%%%%%%%%%%%	Data Model
%%%%%%%%%%%%%%%%%%%%%%%%%%%%%%%%%%%%%%%%%%%%%%%%%%%%%%%%
%%%%%%%%%%%%%%%%%%%%%%%%%%%%%%%%%%%%%%%%%%%%%%%%%%%%%%%%	Data Model
%%%%%%%%%%%%%%%%%%%%%%%%%%%%%%%%%%%%%%%%%%%%%%%%%%%%%%%%

%%%%%%%%%%%%%%%%%%%%%%%%%%%%%%%%%%%%%%%%%%%%%%%%%%%%%%%%	Data Model
%%%%%%%%%%%%%%%%%%%%%%%%%%%%%%%%%%%%%%%%%%%%%%%%%%%%%%%%

%%%%%%%%%%%%%%%%%%%%%%%%%%%%%%%%%%%%%%%%%%%%%%%%%%%%%%%%	
%%%%%%%%%%%%%%%%%%%%%%%%%%%%%%%%%%%%%%%%%%%%%%%%%%%%%%%%	
%%%%%%%%%%%%%%%%%%%%%%%%%%%%%%%%%%%%%%%%%%%%%%%%%%%%%%%%	
%%%%%%%%%%%%%%%%%%%%%%%%%%%%%%%%%%%%%%%%%%%%%%%%%%%%%%%%
\setcounter{figure}{0}
\makeatletter 
\renewcommand{\thefigure}{A\@arabic\c@figure}
\makeatother
%%%%%%%%%%%%%%%%%%%%%%%%%%%%%%%%%%%%%%%%%%%%%%%%%%%%%%%%
\begin{figure}
\epsscale{1.0}
\plotone{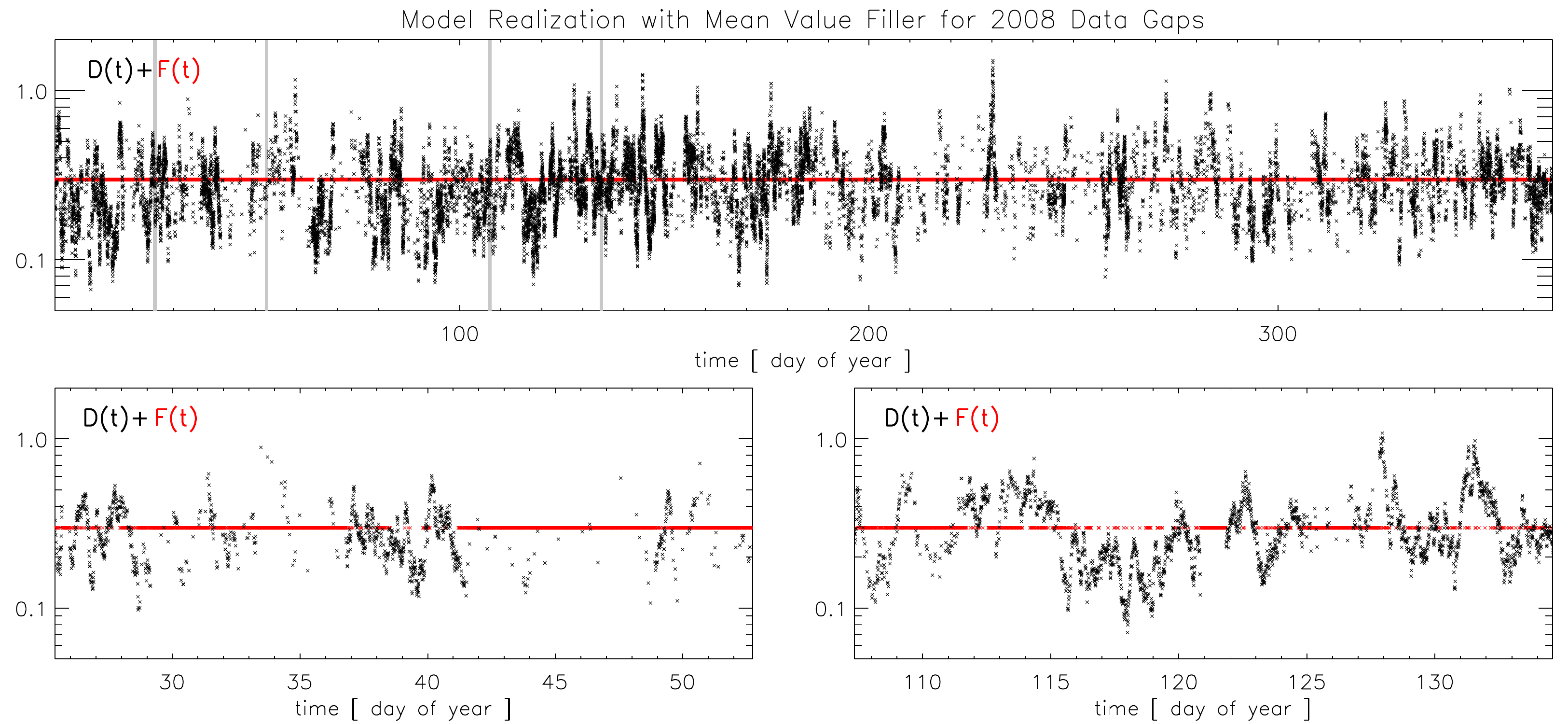}
\caption{The top row plots the synthetic model data time series $D(t)$ used as the example in Section~\ref{S:DataReductionScheme} but with the constant Mean Value filler signal $F(t)$ applied to the observed 2008 data gaps. The bottom row show the two illustrative Carrington Rotation-length periods for details (cf. Figure~\ref{O7O6model}).
}
               \label{fA1}
\end{figure}
%%%%%%%%%%%%%%%%%%%%%%%%%%%%%%%%%%%%%%%%%%%%%%%%%%%%%%%%
\begin{figure}
\epsscale{0.80}
\plotone{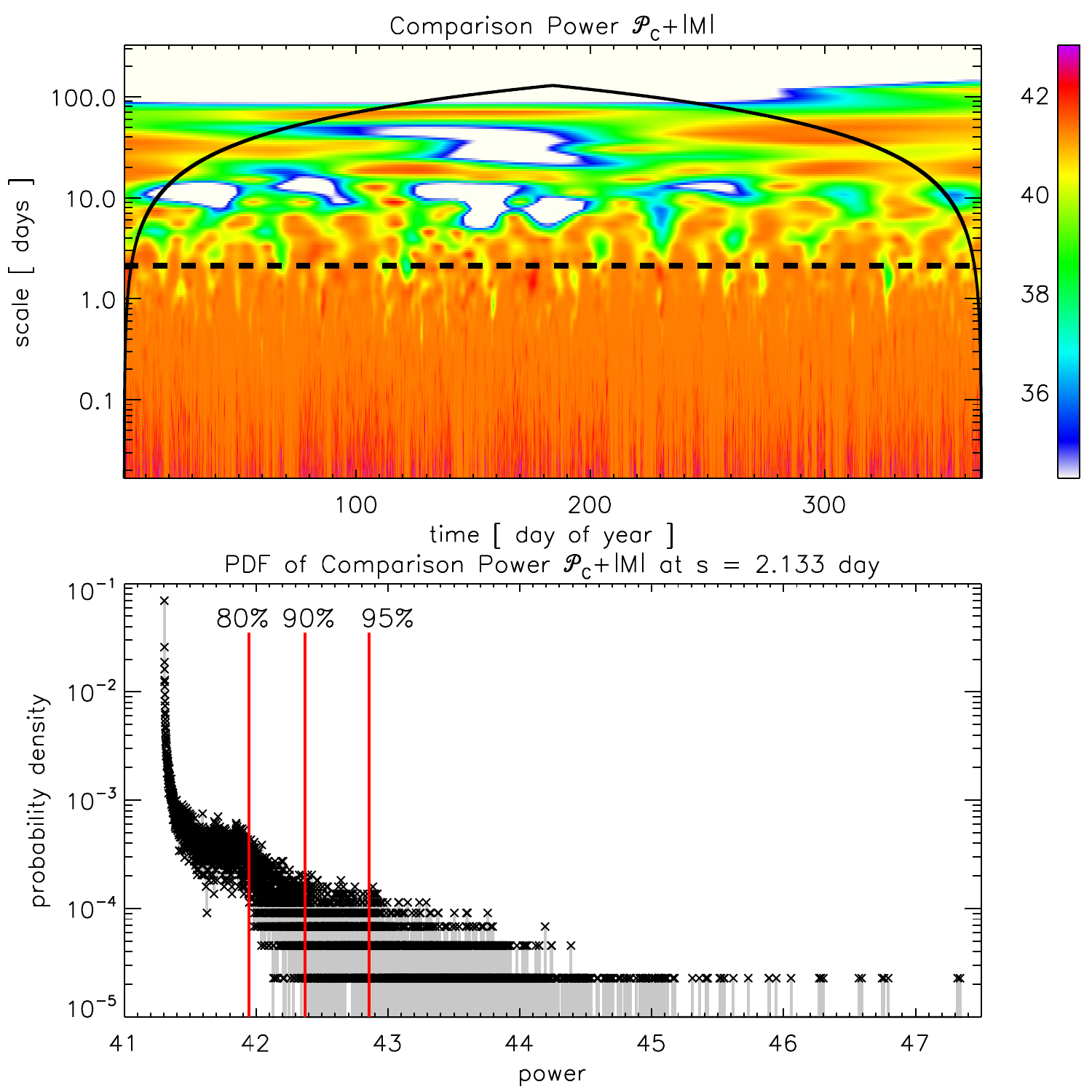}
\caption{Top panel plots the adjusted comparison power spectrum $\mathcal{P}_{C} + |M|$ constructed from the Mean Value filler wavelet power and the resulting interference power. The bottom panel plots the PDF of the comparison power values at the fixed $s=2.133$~day scale with the 80\%, 90\%, and 95\% significance levels also shown (cf. Figure~\ref{WaveletPowerDecomposition2}).
}
               \label{fA2}
\end{figure}
%%%%%%%%%%%%%%%%%%%%%%%%%%%%%%%%%%%%%%%%%%%%%%%%%%%%%%%%
\begin{figure}
\epsscale{0.80}
\plotone{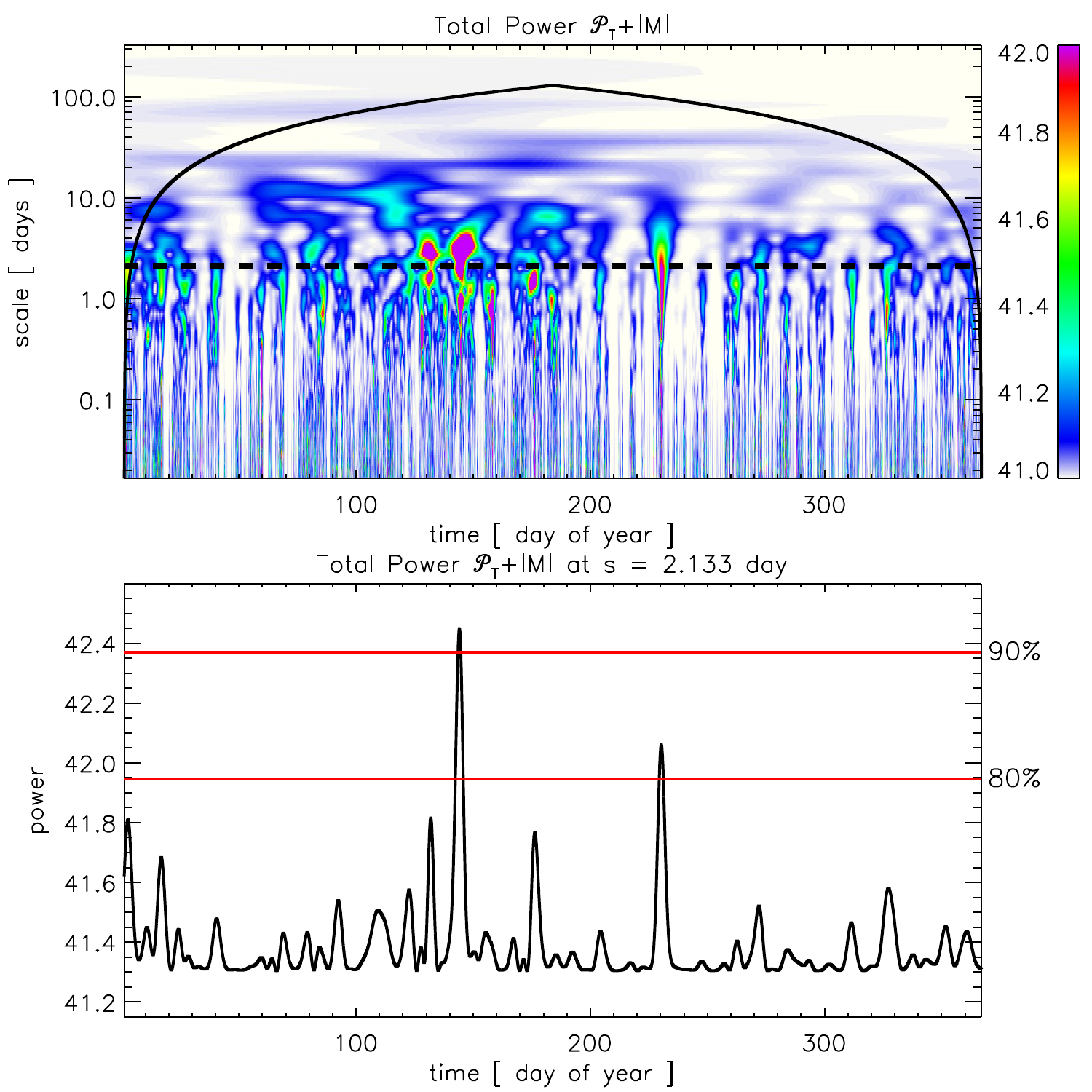}
\caption{Top panel plots the adjusted total power spectrum, $\mathcal{P}_{T} + |M|$ for the synthetic data plus Mean Value filler signal of Figure~\ref{fA1}. Bottom panel plots the total power as a function of time at the $s=2.133$~day scale. Here, the 80\% and 90\% significance levels are shown. Note that none of the total adjusted power at this scale exceeds the 95\% significance level defined by the comparison power PDF (cf. Figure~\ref{WaveletPowerDecomposition3}).
}
               \label{fA3}
\end{figure}

%% End of file %%%%%%%%%%%%%%%%%%
\end{document}